\documentclass[prd,nofootinbib,preprint,superscriptaddress,twocolumn,10pt]{revtex4}
\usepackage{amsmath,amssymb}
\usepackage{soul}
\usepackage{epsfig}
\usepackage{graphicx}
\usepackage[usenames,dvipsnames]{color}
\usepackage{slashed}
\usepackage{ulem,lipsum}
\usepackage[colorlinks,citecolor=blue]{hyperref}
\usepackage{color}
\usepackage{orcidlink}

\def\beq{\begin{equation}}
\def\eeq{\end{equation}}
\def\bey{\begin{eqnarray}}
\def\eey{\end{eqnarray}}

\def\lsim{\mathrel{\raise.3ex\hbox{$<$\kern-.75em\lower1ex\hbox{$\sim$}}}}
\def\gsim{\mathrel{\raise.3ex\hbox{$>$\kern-.75em\lower1ex\hbox{$\sim$}}}}

\def\beq{\begin{equation}\begin{aligned}}
\def\eeq{\end{aligned}\end{equation}}

\newcommand{\bmt}{\begin{pmatrix}}
\newcommand{\itt}{\end{pmatrix}}
\newcommand{\ba}{\begin{array}{c}}
\newcommand{\ea}{\end{array}}
\newcommand{\be}{\begin{equation}}
\newcommand{\ee}{\end{equation}}
\newcommand{\bea}{\begin{eqnarray}}
\newcommand{\eea}{\end{eqnarray}}

\newcommand{\bi}{\begin{itemize}}
\newcommand{\ei}{\end{itemize}}

\newcommand{\baz}{\begin{array}{cc}}

\newcommand{\mathsym}[1]{{}}

\newcommand{\bt}{\begin{tabular}}
\newcommand{\et}{\end{tabular}}

\newcommand{\benu}{\begin{enumerate}}
\newcommand{\eenu}{\end{enumerate}}
\newcommand{\bav}{\begin{array}{cccc}}

\begin{document}
\title{\bf A Common Origin of Asymmetric Self-interacting Dark Matter and Dirac Leptogenesis}
\author{Manoranjan Dutta\orcidlink{0000-0003-3425-7050}}
\email{md@nlu.ac.in}
\affiliation{Department of Physics, North Lakhimpur University, Khelmati, North Lakhimpur, Lakhimpur, Assam 787031, India}

\author{Nimmala Narendra\orcidlink{0000-0002-8577-0324}}
\email{nimmalanarendra@gmail.com}
\affiliation{Department of Physics, PES Institute of Technology \& Management,
Sagar Road, Shivamogga, Karnataka-577204, India}



\begin{abstract}
Assuming dark matter to be asymmetric as well as self-interacting and neutrinos to be Dirac fermions, we propose a framework to address the observed baryon imbalance of the universe.  We add three right-handed neutrinos $\nu_{R_i},\,{i=1,2,3}$, one singlet fermion $\chi$, a doublet fermion $\psi$, and heavy scalar doublets $\eta_i,\,{i=1,2}$ to the Standard Model. A global $B-L$ is imposed to protect the Dirac nature of neutrinos. Both $\chi$ and $\psi$ are fermions with non-zero charge under an extended $U(1)_{B-L} \times U(1)_D$ symmetry. Additionally, a $\mathcal{Z}_2$ symmetry is imposed, where the singlets $\chi$, $\nu_R$, and $\eta$ are negative and the doublet $\psi$ is positive.  The CP-violating out-of-equilibrium decay of heavy scalar $\eta$ generates an equal and opposite $B-L$ asymmetry among the left-handed ($\nu_L$) and right-handed ($\nu_R$) neutrinos. The $\nu_L-\nu_R$ equilibration process does not take place until below the Electroweak phase transition scale because of tiny Yukawa couplings. During this time, Sphaleron processes, which are active at temperatures higher than 100 GeV, transform a portion of the $B-L$ asymmetry stored in left-handed neutrinos into baryon asymmetry. MeV scale gauge boson $Z'$ of $U(1)_D$ sector mediates both annihilation of symmetric dark matter component and self-interaction among dark matter particles. Moreover, $Z'$ mixes with the Standard Model Z-boson and provides a portal for dark matter direct detection.
\end{abstract}

\maketitle

\newpage

\section{Introduction} \label{Intro}
The Standard Model (SM) is a remarkably successful theory explaining interactions of elementary particles. Yet several mysteries remain unanswered in SM, of which the non-zero neutrino mass, the observed baryon asymmetry of the Universe, and the candidature problem of dark matter (DM), {\it etc.}, are the most prominent issues. The rotation velocities of stars in galaxies, gravitational lensing, anisotropies in the CMB, {\it etc.}, hint the existence of an electromagnetically inert form of matter. This new form of matter is named dark matter (DM). The particle nature of DM is yet to be revealed. WMAP~\cite{Hinshaw:2012aka} and PLANCK~\cite{Aghanim:2018eyx} data (at 68\% CL) quantify the relic baryon and DM density as:
 $\Omega_{\rm B} h^{2}=0.02237 \pm 0.00015$, $\Omega_{\rm DM} h^{2}=0.1200 \pm 0.0012$;
 where $\Omega_{\rm DM}$ is the ratio of DM density to the critical density of the Universe and $h = \text{Hubble Parameter}/(100 \;\text{km} ~\text{s}^{-1} 
\text{Mpc}^{-1})$ is the reduced Hubble constant. Observations also indicate that the present Universe is highly baryon asymmetric. The observed Baryon Asymmetry of the Universe (BAU) can be  quantified as~\cite{Aghanim:2018eyx}, 
\begin{equation}
\eta_B = \frac{n_{B}-n_{\overline{B}}}{n_{\gamma}} \simeq 6.2 \times 10^{-10},
\label{etaBobs}
\end{equation}
where, $n_{B}$, $n_{\overline{B}}$ and $n_\gamma$ are the number densities of baryon, antibaryon and photon respectively. Based on Sakharov conditions~\cite{Sakharov:1967dj,sakharov.67}, the observed baryon asymmetry has been addressed in frameworks of leptogenesis~\cite{fukugita.86,baryo_lepto_group_1,baryo_lepto_group_2,baryo_lepto_group_3,
baryo_lepto_group_4,baryo_lepto_group_5,baryo_lepto_group_6,baryo_lepto_group_7,baryo_lepto_group_8,baryo_lepto_group_9,Abada:2006ea,Pilaftsis:2003gt,Pilaftsis:2005rv,Kaplan:2009ag,Feng:2012jn,Kolb:1990vq,Ibe:2011hq,Feng:2013wn,Narendra:2018vfw}. WMAP and Planck data imply that $\Omega_{\rm DM} \approx 5 \Omega_{\rm B}$. It triggers to assume DM to have a similar origin as that of baryons and address both phenomenologies in a common framework~\cite{old_DM_Baryon_asy_1, old_DM_Baryon_asy_2, old_DM_Baryon_asy_3, old_DM_Baryon_asy_4, old_DM_Baryon_asy_5,
old_DM_Baryon_asy_6,old_DM_Baryon_asy_7,old_DM_Baryon_asy_8, Asydm_models1_1,Asydm_models1_2,Asydm_models1_3,Asydm_models1_4,
Asydm_models1_5,Asydm_models1_6,Asydm_models1_7,Asydm_models1_8,Asydm_models1_9,Asydm_models1_10, Asydm_models2_1,Asydm_models2_2,
Asydm_models2_3,Asydm_models2_4,Asydm_models2_5,Asydm_models2_6,Asydm_models2_7,Asydm_models2_8,Asydm_models2_9,Arina:2011cu,Arina:2012fb,Arina:2012aj,Narendra:2017uxl,Asydm_review_1,Asydm_review_2,Kanemura:2025ixp}.

From cosmological standpoint, DM is assumed to be a cold, collisionless fluid by the ${\rm \Lambda CDM}$ model. ${\rm \Lambda CDM}$ model has proved incredibly successful in explaining the large-scale structures of the Universe. Nevertheless, a number of discrepancies have surfaced at smaller scales, such as the core-cusp problem, the missing satellite problem, and the too-big-to-fail problem {\it etc}~\cite{Tulin:2017ara, Bullock:2017xww}. These discrepancies are more noticeable at the dwarf galaxy scale and progressively diminish to the cluster scale, which corresponds with $\Lambda{\rm CDM}$ predictions at large scales. In order to mitigate these anomalies, self-interacting dark matter (SIDM)~\cite{Carlson:1992fn, deLaix:1995vi,Spergel:1999mh} was suggested as a substitute for traditional cold dark matter. A velocity-dependent self-scattering of order $\sigma/m \sim 1 \; {\rm cm}^2/{\rm g} \approx 2 \times 10^{-24} \; {\rm cm}^2/{\rm GeV}$  has been demonstrated to alleviate these anomalies~\cite{Buckley:2009in, Feng:2009hw, Feng:2009mn, Loeb:2010gj, Zavala:2012us, Vogelsberger:2012ku, Bringmann:2016din, Kaplinghat:2015aga, Kamada:2020buc, Aarssen:2012fx, Tulin:2013teo}. DM models with a light mediator is well suited to realise such velocity dependence~\cite{Kouvaris:2014uoa, Bernal:2015ova, Kainulainen:2015sva, Hambye:2019tjt, Cirelli:2016rnw, Kahlhoefer:2017umn, Ghosh:2021wrk,Yang:2025dgl}. However a common issue with SIDM is its under-abundant relic due to efficient annihilation into the light mediator \cite{Dutta:2021wbn,Borah:2021yek,Borah:2021pet,Borah:2021rbx,Dutta:2024ydz}. However, if SIDM is assumed to have similar origin as that of baryons, correct relic density can be achieved~\cite{Dutta:2022knf, Borah:2024wos}.

Because there are no right-handed neutrino fields, neutrinos have no mass within the SM framework. However, neutrino flavor oscillation~\cite{Abe:2011sj, Abe:2013hdq,Abe:2011fz, DoubleChooz:2019qbj, An:2012eh, An:2017osx, Adey:2018zwh,Ahn:2012nd,Adamson:2013whj, Adamson:2014vgd} suggests that the neutrino mass is non zero but small. Beyond the SM framework invoking lepton number violating dimension-5 operator $LLHH/\Lambda$~\cite{Weinberg:1979sa} is a lucrative method of explaining the small mass of Majorana neutrinos. The scale of new physics is $\Lambda$, whereas the lepton and Higgs doublets are denoted by $L, H$. In  this way, neutrinos obtain a Majorana mass of order $m_\nu = \langle H \rangle^2/\Lambda$ during electroweak phase transition (EWPT). This novel framework is referred to as {\it seesaw mechanism} includes type-I~\cite{Minkowski:1977sc, Yanagida:1981xy, GellMann:1980vs, Mohapatra:1979ia}, type-II~\cite{type_II_seesaw_1, type_II_seesaw_2, type_II_seesaw_3, type_II_seesaw_4, type_II_seesaw_5,type_II_seesaw_6,
type_II_seesaw_7}, type-III~\cite{type_III_seesaw} or their variants~\cite{ray_volkas_review}. This mechanism works well for $\Lambda \sim {\cal O}(10^{14}){\rm GeV}$ for neutrino masses below eV. Neutrinoless double beta decay experiments indicative of the Majorana nature of neutrinos~\cite{onu_beta_expt} are  being investigated. However, as of yet, no corroborating results have been found. Consequently, there is still a chance that neutrinos could be Dirac fermions.  

Considering that neutrinos are Dirac particles (which requires $B-L$ to be an exact symmetry), Dirac Leptogenesis can account for the observed baryon asymmetry of the Universe~\cite{Dirac_leptogenesis_1,Dirac_leptogenesis_2,Dirac_leptogenesis_3,Dirac_leptogenesis_4,
Dirac_leptogenesis_5,Dirac_leptogenesis_6,Dirac_leptogenesis_7}. The fundamental mechanism of Dirac Leptogenesis is that the equilibration time between left and right-handed neutrinos mediated via Yukawa interaction with SM Higgs (i.e. $y\overline{\nu_R}H\nu_L$) is much less than the $(B+L)$ violating sphaleron transitions above electroweak phase transition. Therefore, demanding $B-L= B-(L_{\rm SM}+L_{\nu_R})=0$\cite{Cerdeno:2006ha}, a net $B-L_{\rm SM}$ can be generated in terms of $L_{\nu_R}$. $\nu_R$ being $SU(2)_L$ singlet, $L_{\nu_R}$ remains unaffected by sphalerons. The non-zero $B-L_{\rm SM}$ is, however, converted to a net $B$ asymmetry via $B+L$ violating sphaleron transitions~\cite{Dick,Buchmuller:2000as,Cerdeno:2006ha,Gu:2006dc,Gu:2007mc,Murayama:2002je,Thomas:2005rs,Borah:2016zbd,Perez:2021udy}. 

\begin{figure} [h!]
\centering
\includegraphics[width=60mm]{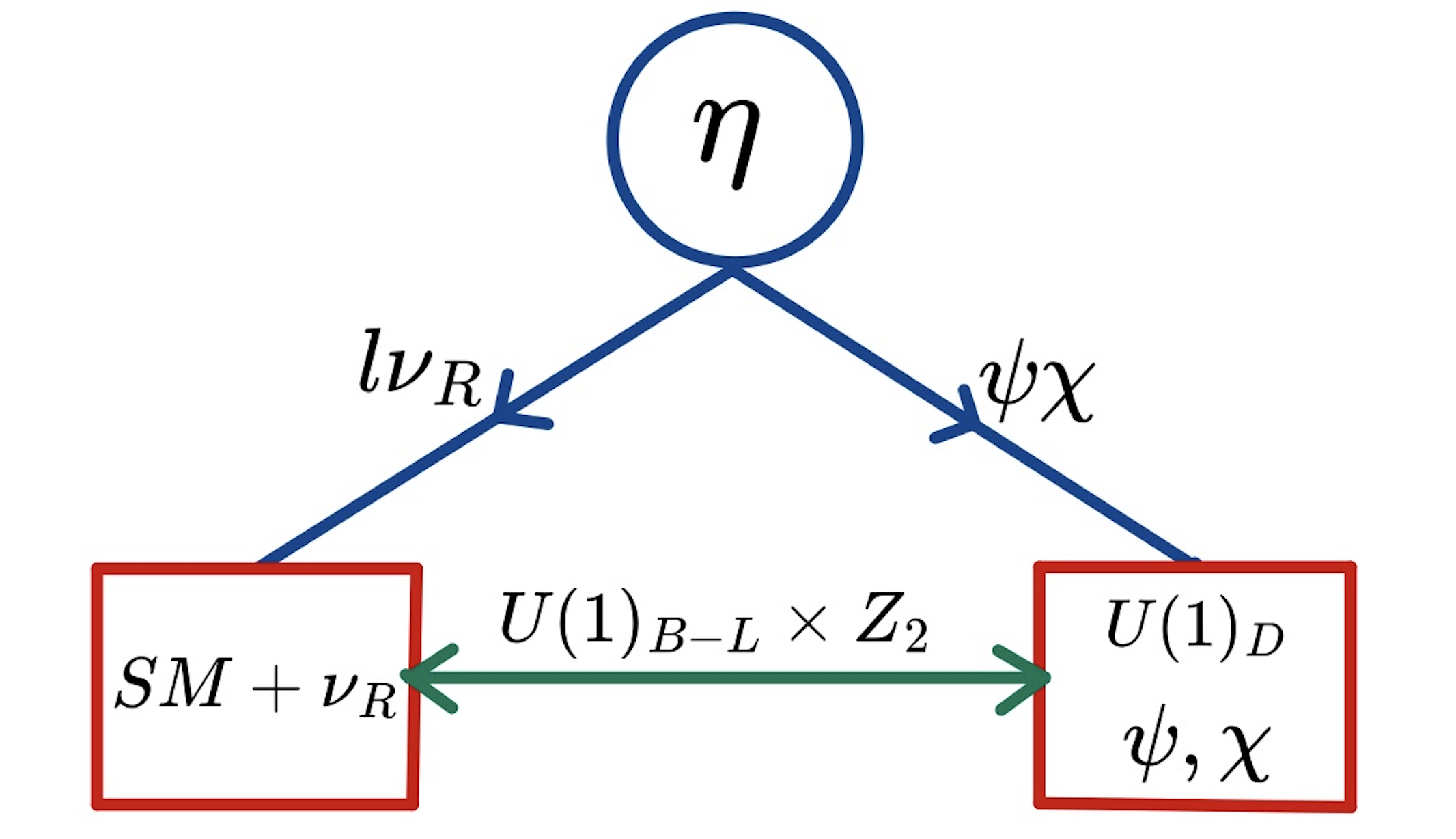}
\caption{Schematic of the model: A heavy scalar $\eta$ decays into visible ($l\nu_R$) and dark sectors ($\psi \chi$), where the dark sector includes the particles charged under $U(1)_D$ symmetry. The lightest dark sector particle $\chi$ is the DM candidate.}
\label{Diagram}
\end{figure}

To integrate all these avenues, in this paper, we construct a framework for explaining DM and baryon asymmetry considering Dirac neutrinos. The asymmetry in both sectors is generated by the CP-violating out-of-equilibrium decay of a heavy scalar doublet $\eta$ into  $\ell \nu_R$ and $\psi \chi$, where $\ell$ is SM lepton, $\nu_R$ is right handed neutrino (RHN), and $\psi$, $\chi$ are respectively a vector-like fermion doublet and a vector-like fermion singlet that comprises the dark sector. The lepton asymmetry in the channel $\eta \to \ell \nu_R$ is then converted to a net baryon asymmetry via $B+L$ violating sphaleron transitions. Being a vector-like Dirac fermion, asymmetry in the channel $\eta \to \psi \chi$ remains intact~\cite{Arina:2011cu,Arina:2012fb,Arina:2012aj,Narendra:2017uxl}. The dark sector is gauged under a $U(1)_D$ symmetry. The corresponding gauge boson $Z'$ mediates DM self interaction. The lightest dark sector particle $\chi$ which is odd under $\mathcal{Z}_2$ acts as the SIDM candidate. The symmetric DM component annihilates away into a pair of $Z'$, yielding a vanishingly small symmetric relic. We also impose $U(1)_{B-L}\times \mathcal{Z}_{2}$ symmetry which ensures that the neutrinos are Dirac. $\mathcal{Z}_2$ symmetry is softly broken by a term $\mu^2 \eta^\dagger H$, generating  Dirac mass for neutrinos.  A schematic of the framework is shown in Fig.~\ref{Diagram}.

The paper is organized as follows. In sec.~\ref{The Model}, we discuss the proposed model, while in sec.~\ref{Dirac neutrino mass}, we explain the Dirac mass of light neutrinos.
Section \ref{Production of LA} is devoted to explain baryogenesis via leptogenesis from the decay of heavy particles $\eta$, while section \ref{Dark matter} 
describes the mechanism for DM relic abundance. The phenomenology of SIDM is discussed in section \ref{sidm}, followed by constraints from direct search in sec.~\ref{sidm_dd}. Then we discuss possible collider phenomenology in sec.~\ref{coll}. We finally conclude in sec.~\ref{Conclusion}.

\section{The Model}\label{The Model}
We extend the SM both by new symmetries and particles. The new particles include: three RHNs $\nu_{{R}_{\alpha}}$, $ \alpha = 1,2,3$, a heavy scalar doublet $\eta$, and a pair of vector-like leptons, one singlet ($\chi$) and the other doublet ($\psi$) under $SU(2)$. The particle content beyond those in the SM and their charge assignments under the extended symmetries are shown in Table~\ref{tab1}.

\begin{table}[!h]
\renewcommand{\arraystretch}{}
\centering
\begin{tabular}{|c|c|c|c|}
\hline
Parameter             & $U(1)_{B-L}$ & $U(1)_{D}$ & $\mathcal{Z}_{2}$   \\
\hline
$\eta=(\eta^{+},\eta^{0})^{T}$ &       0      &      0     &    -      \\
\hline
$\nu_{R}$             &      -1      &      0     &    -      \\
\hline
$\psi=(\psi^{0},\psi^{-})^{T}$ & -1  &      1     &    +      \\
\hline
$\chi$                &      -1      &      1     &    -      \\
\hline
\end{tabular}
\caption{Quantum numbers of the new particles under the extended symmetry.}
\label{tab1}
\end{table}

A discrete symmetry $\mathcal{Z}_2$ is introduced under which $\eta,\nu_{R}$ and $\chi$ are odd, while $\psi$ and all other SM particles are even. The fermions $\psi$ and $\chi$ also carry quantum numbers under $U(1)_{D}$ symmetry. $\chi$ is the lightest of dark sector particles, hence the DM candidate. $U(1)_{D}$ symmetry is spontaneously broken and MeV scale $U(1)_D$ gauge boson $Z'$ mediates DM self-interaction. A global $U(1)_{B-L}$ symmetry is imposed so that only Dirac mass term is
allowed for neutrinos, while Majorana mass terms are forbidden, ensuring lepton
number conservation at the level of the renormalisable Lagrangian.
Owing to charge assignments of the fields as in Table~\ref{tab1}, the relevant Lagrangian for the phenomenologies is given by: 
\begin{equation}
 \begin{split}
        \mathcal{L} & \supset \bar{\psi}i\gamma^{\mu} D_{\mu}\psi + \bar{\chi}i \gamma^{\mu}D'_{\mu}\chi + M_{\psi}\bar{\psi} \psi + M_\chi \bar{\chi} \chi  \\
        & + \left[ f_{kl} \overline{\ell_{k}} \tilde{\eta} {\nu_{R}}_{l} + \lambda \overline{\psi} \tilde{\eta} \chi + {\rm h.c.}\right] - V(H, \eta) \,,
    \end{split}
    \label{Eq-Lagrangian}
\end{equation}
where $\tilde{\eta} = i \tau_2 \eta^*$, $\tau_2$ being the second Pauli spin matrix and,
\begin{eqnarray}
D_{\mu} &=& \partial_{\mu}-i\frac{g}{2}\tau^{i}W^{i}_{\mu}-i\frac{g_{_Y}}{2}YB_{\mu}-ig' Y' (Z')_{\mu},
 \nonumber\\
D'_{\mu} &=& \partial_{\mu}-ig' Y' (Z')_{\mu},
\end{eqnarray}
and the scalar potential is given by,
\begin{eqnarray}
V(H, \eta) &=& -M^{2}_{H} H^{\dagger} H + \lambda_{H} (H^{\dagger} H)^{2} \nonumber\\
&+& M_\eta^2 \eta^{\dagger} \eta + \lambda_\eta (\eta^{\dagger} \eta)^{2} + \lambda_{H \eta} (H^{\dagger} H) (\eta^\dagger \eta). 
\label{potential}
\end{eqnarray}

Heavy scalar $\eta$ undergoes CP-violating out-of-equilibrium decay. This decay creates asymmetries in both lepton and dark sectors simultaneously~\cite{Arina:2011cu}. The process $\eta \to \ell \nu_R$, creates equal and opposite lepton asymmetry in left and right-handed sectors. $B+L$ violating sphaleron transitions converts the lepton asymmetry in the left-handed sector to a net baryon asymmetry. Asymmetry in the right-handed sector remains unaffected until the temperature falls much below the electroweak phase transition. DM $\chi$ attains thermal equilibrium in early universe via processes like $\psi \psi \to \chi \chi$, mediated by the $U(1)_D$ gauge boson $Z'$. However, $\chi$ annihilates efficiently into pairs of $Z'$ via t-channel and a negligible symmetric DM component is left. The correct DM relic density is obtained from the CP-violating out-of-equilibrium decay of heavy scalar $\eta$. Hence, in this model, the observed DM relic is dominantly asymmetric. Kinetic mixing between $Z'$ and SM $Z$ enables direct detection of DM in terrestrial laboratories.
$Z'$ also facilitates DM self-interaction. The term $\mu^2 \eta^\dagger H$ breaks $\mathcal{Z}_2$ symmetry softly generating Dirac mass for neutrinos. The global $B-L$ symmetry ensures Dirac nature of the neutrinos forbidding Majorana mass. 

However, it should be noted that, global symmetries are generically expected to be violated by
quantum-gravitational effects~\cite{Banks:2010zn}. Consequently, a global
$U(1)_{B-L}$, even if imposed at the renormalisable level, may be broken
by Planck-suppressed operators that reintroduce Majorana mass terms.
For this reason, the theoretically robust way to protect Dirac neutrino
masses is to gauge the $U(1)_{B-L}$ symmetry.  In a gauged theory,
Majorana mass terms are forbidden by gauge invariance at all orders, and
quantum-gravity effects cannot violate the symmetry since possible
Planck-suppressed operators are tightly constrained by gauge charge
assignments.  With the charges listed in Table~\ref{tab1}, our Lagrangian is
already invariant under $U(1)_{B-L}$, and the field content (SM plus
three $\nu_R$) ensures anomaly cancellation.  Dark fermions $\psi$
and $\chi$ being vectorlike, do not contribute
to gauge anomalies. Thus, the model admits a consistent gauged $B-L$
embedding without altering interactions
relevant to the Dirac leptogenesis mechanism.
Gauging $B-L$ simply introduces the corresponding gauge boson together
with the usual kinetic term and covariant derivatives, plus an optional
$B-L$–breaking scalar $\phi_{B-L}$ whose $B-L$ charge can be chosen so as not
to generate the unwanted operator $\phi_{B-L}\nu_R\nu_R$, thereby
preserving the Dirac nature of neutrinos.  If the $B-L$ breaking scale
is sufficiently high ($M_{Z_{B-L}}\gg\mathrm{TeV}$), the new gauge boson
effectively decouples, leaving the low-energy phenomenology of our model
unchanged.  In this case, the primary role of gauged $B-L$ is to protect
the Dirac neutrino mass from quantum-gravity violations of global
symmetries.

\section{Dirac mass of neutrinos}\label{Dirac neutrino mass}
\begin{figure} [h!]
\centering
\includegraphics[width=50mm]{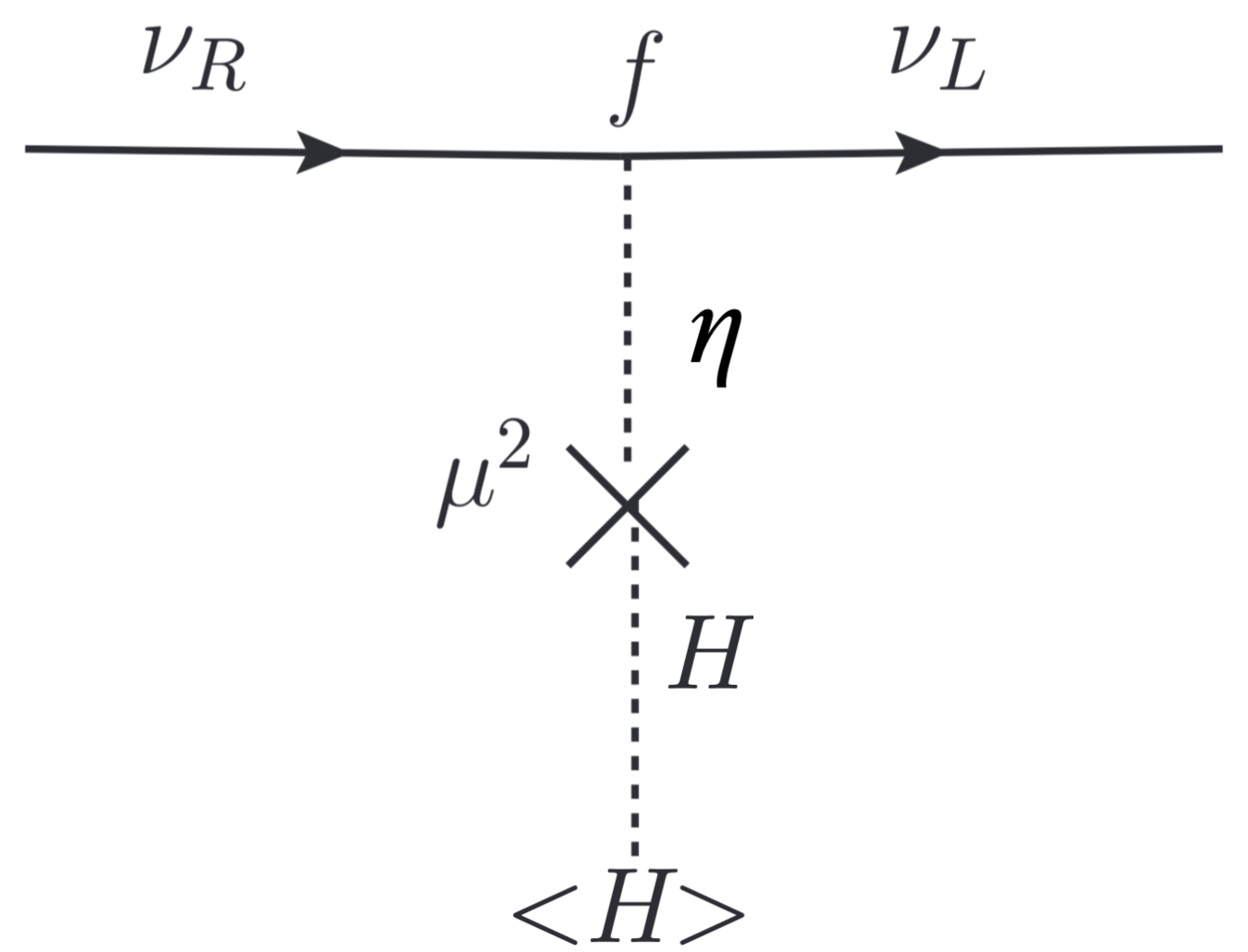}
\caption{Feynman diagram responsible for generation of Dirac neutrino mass.}
\label{Dirac neutrino}
\end{figure}

Imposed $\mathcal{Z}_2$ symmetry is softly broken~\cite{McDonald:2007ka, Heeck:2013vha} by the Lagrangian  term:
\begin{equation}
\mathcal{L}_{soft} = -\mu^{2} H^{\dagger}\eta + {\rm h.c.}
\end{equation}
Therefore, Dirac mass of the neutrinos can be generated as shown in the Fig.~\ref{Dirac neutrino}.
After integrating out the heavy field $\eta$, the Dirac mass is obtained as,
\begin{equation}
\label{nu}
M_\nu = \frac{f \langle H \rangle \mu^{2}}{M_\eta^{2}}\,,
\end{equation}
where $\langle H \rangle = 174$ GeV, is the vacuum expectation value of the SM Higgs. From Eq.~\ref{nu}, we get neutrino mass of order $0.1$ eV, for $\frac{\mu}{M_\eta} \approx 10^{-4} $ and $f \sim 10^{-4}$. The smallness of $\mu$ compared to $M_\eta$ justifies the soft breaking.

\section{Baryogenesis via Dirac Leptogenesis from $\eta$ decay}\label{Production of LA}
Initially the heavy scalar $\eta$ is in thermal equilibrium due to its gauge and Yukawa interactions at a temperature above its mass scale. As the Universe expands and cools down, the heavy scalar goes out-of-equilibrium and decays at a temperature $T \sim M_\eta$. We consider two copies of the scalar $\eta$ to achieve the CP-violation. The CP-violating out-of-equilibrium decay generates an asymmetry in both visible ($\ell \nu_R$) and dark sectors ($\psi \chi$). The decay width of the $\eta$ particle can be written as,
\begin{equation}
    \Gamma_\eta \simeq \frac{1}{8 \pi} (f^2+\lambda^2) M_\eta,
\end{equation}
where $f$ and $\lambda$ are the Yukawa couplings. In presence of pairs of $\eta$-particles and their interactions, the diagonal mass terms for $\eta$ in Eq.~\ref{potential} can be replaced by~\cite{Ma:1998dx,Arina:2011cu},
\begin{equation}
\frac{1}{2} \eta^{\dagger}_{a}(M_{+}^{2})_{ab}\eta_{b}+\frac{1}{2} (\eta^{*}_{a})^{\dagger}(M_{-}^{2})_{ab}\eta^{*}_{b}\,,
\end{equation}
Where
\begin{equation}
M^{2}_{\pm}= 
\begin{pmatrix} M_{1}^{2}-i C_{11} & -i C_{12}^{\pm}\cr\\
-i C_{21}^{\pm} & M_{2}^{2}-i C_{22} \end{pmatrix}
\label{mass_matrix},
\end{equation}
here
$C_{ab}^{+}=\Gamma_{ab}M_{b}$, $C_{ab}^{-}=\Gamma^{*}_{ab}M_{b}$ and $C_{aa}=\Gamma_{aa}M_{a}$ with
\begin{equation}
\Gamma_{ab}M_{b}=\frac{1}{8\pi}\left(M_{a}M_{b}\lambda_{a}\lambda_{b}^{*}+M_{a}M_{b} \mathop{\sum_{k,l}} f_{akl}^{*}f_{bkl}\right).
\end{equation}

\begin{figure} [h!]
\centering
\includegraphics[width=90mm]{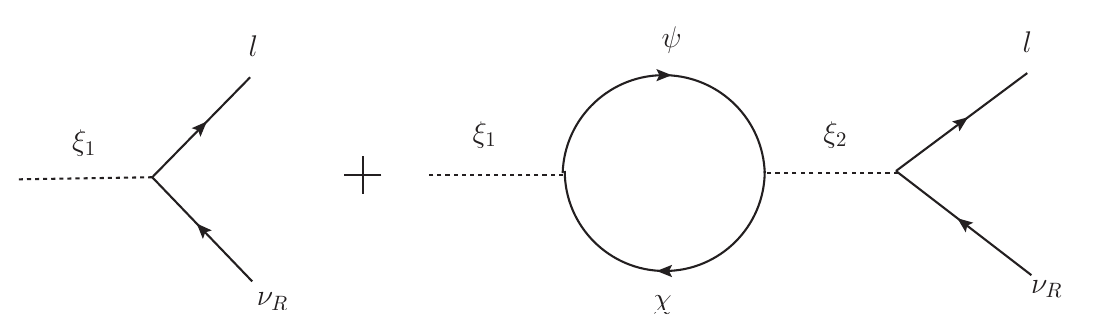}
\caption{The interference of Tree level and Self-energy correction diagrams that contribute to the CP violation.}
\label{LA}
\end{figure}

After diagonalizing the mass matrix given in Eq.~\ref{mass_matrix}, we get two mass eigenstates $\xi_1^{\pm}$ and $\xi_2^{\pm}$ with two mas eigenvalues $M_{\xi_1}$ and $M_{\xi_2}$. Their decay provide CP-asymmetry via the interference of tree level and one-loop self-energy correction diagrams, shown in Fig.~\ref{LA}\footnote{Note that the mass eigenstates $\xi_1^+$ and $\xi_1^-$ (similarly $\xi_2^+$ and $\xi_2^-$) are not CP conjugates of each other even though they are degenerate mass eigenstates.}. The generated asymmetry in the visible sector can be quantified as,
\begin{eqnarray}
\epsilon_{L} &=& [B_L(\xi_{1}^{-} \rightarrow l^{-} \nu_{R})-B_L(\xi_{1}^{+}\rightarrow (l^{-})^{c} \nu_{R}^{c})]\nonumber\\
  &=& -\frac{Im\left(\lambda_{1}^{*} \lambda_{2} \mathop{\sum_{k,l}} f^{*}_{1kl} f_{2kl} \right)}{8\pi^2(M_{2}^{2}-M_{1}^{2})} \left[\frac{M_{1}^{2}M_{2}}{\Gamma_{1}} \right]\,,
 \end{eqnarray}
where $B_L$ is the branching ratio for $\xi_{1}^{\pm} \rightarrow l^{\pm} \nu_{R}$. We can estimate the generated lepton asymmetry in the visible sector using the relevant Boltzmann equations governing the evolution of the number density of $\xi_1$ and the lepton asymmetry:
\begin{equation}
\begin{split}
\frac{dY_{\xi_{1}}}{dx}&=-\frac{x}{H(M_{\xi_{1}})}s<\sigma|v|_{(\xi_{1}\xi_{1}\rightarrow All)}>\left[ Y_{\xi_{1}}^{2}-{Y_{\xi_{1}}^{eq}}^2 \right]\\
&-\frac{x}{H(M_{\xi_{1}})}\Gamma_{(\xi_{1} \rightarrow All)} \left[ Y_{\xi_{1}}-Y_{\xi_{1}}^{eq} \right]\\
\frac{dY_{L}}{dx}&=\epsilon_{L} \frac{x}{H(M_{\xi_{1}})}\Gamma_{(\xi_{1}\rightarrow All)}B_{L} \left[ Y_{\xi_{1}}-Y_{\xi_{1}}^{eq} \right]\,,
\end{split}
\end{equation}
where the $x=M_{\xi_{1}}/T$. The lepton and $\xi$ abundance are defined as $Y_{\xi, L} \equiv n_{\xi,L}/s$, where $s=(2\pi^2/45)g_* T^3$ is the entropy density of the Universe. 
\begin{figure} [h!]
\centering
\includegraphics[width=90mm]{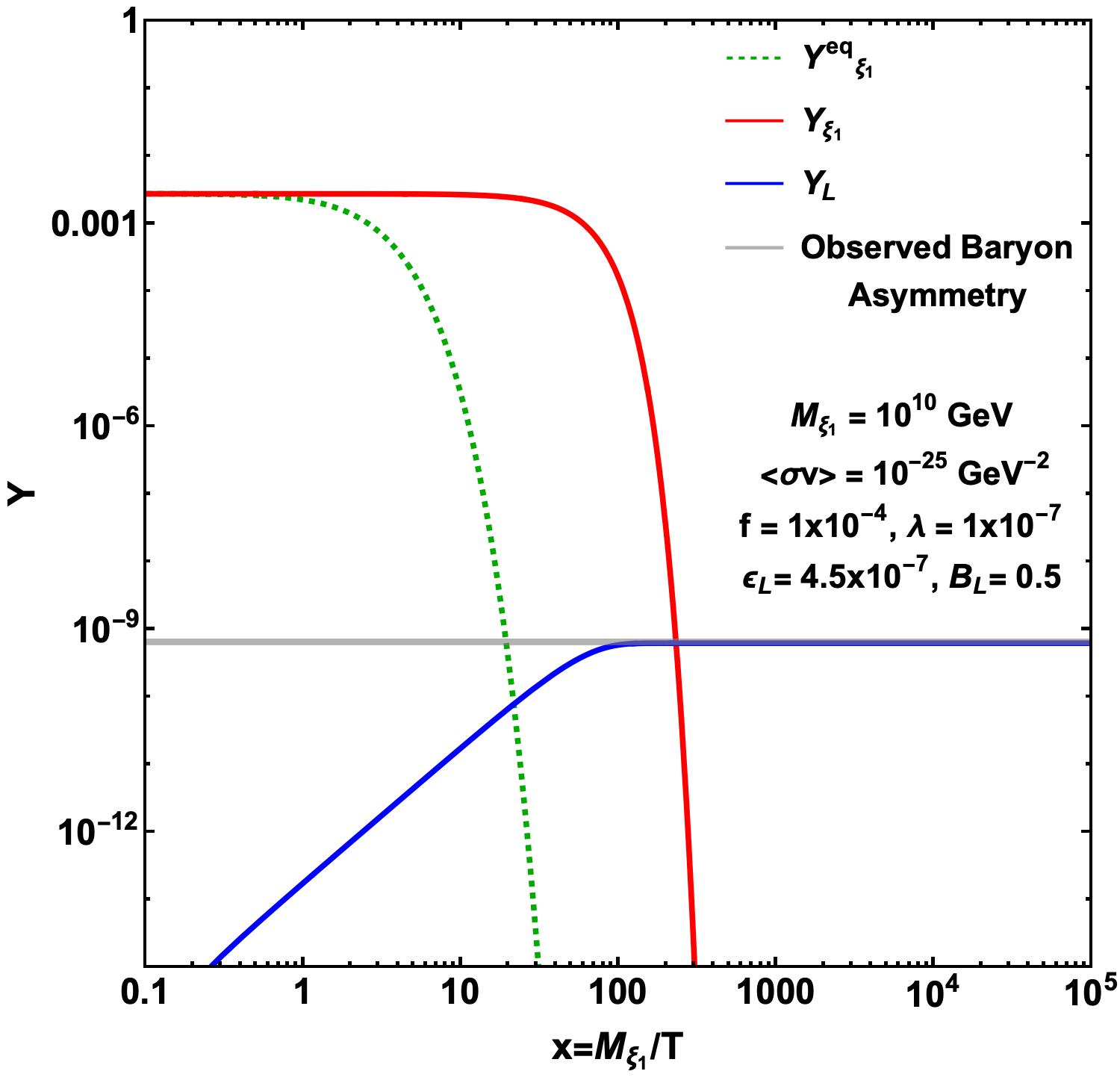}
\caption{The lepton asymmetry from $\xi_1$ decay. The Blue line shows the evolution of lepton asymmetry for $\epsilon_{L} = 4.5 \times 10^{-7}$. The Red line shows the evolution of $\xi_1$ particles. The Green(dotted) line shows the equilibrium number density of $\xi_1$.}
\label{yL_yDM}
\end{figure}

Fig.~\ref{yL_yDM} depicts lepton asymmetry $Y_{L}$ and comoving number density of $\xi_{1}$, {\it i.e.,} $Y_{\xi_{1}}$, as a function of the dimensionless parameter $x$. Coupling constants $f$ is taken as $10^{-4}$ and $\lambda$ is taken as $10^{-7}$. The typical value of the cross-section is taken as $\sigma|v|_{(\xi_{1}\xi_{1} \rightarrow All)}=10^{-25} \,{\rm GeV}^{-2}$. The solid red and blue curves show the evolution of number densities of $\xi_{1}$ and lepton asymmetry respectively. As temperature falls below the mass of $\xi_1$ ({\it i.e.,} $x > 1$), it decouples from the thermal bath (see Fig.~\ref{yL_yDM}). As $\xi_{1}$ starts decaying to $\ell^- \nu_R$, lepton asymmetry develops proportional to $B_L$ and $\epsilon_L$. Once the decay of $\xi_1$ is complete, the lepton asymmetry settles to a constant value. In Fig.~\ref{yL_yDM}, we have taken the branching ratio $B_L = 0.5 $, $\epsilon_{L}$ = $ 5 \times 10^{-7}$ and the heavy scalar mass to be $M_{\xi_{1}}$ = $10^{10}$ GeV. Electroweak Sphalerons that violate $B+L$ (but conserve $B-L$) transfer a portion of the lepton asymmetry to a net baryon asymmetry $Y_{B} = -0.55 Y_{L}$.

Possible washout of the visible lepton asymmetry would require 
(i) chemical equilibration between the dark and visible sectors and 
(ii) an active lepton-number-violating (LNV) process.  In our model, 
neither condition is realised. The only portal after $\eta$ decouples is 
the tiny kinetic mixing $\epsilon\lesssim10^{-8}$, which ensures that, scattering rate of all relevant interactions is too slow compared to the Hubble rate. Besides, once the temperature drops below $T\ll M_{\xi_i}$, there is no operational LNV channel for any dark-visible process to couple to. Hence neither the dark nor the visible asymmetry is exposed to washout. A detailed wash-out analysis is provided in 
Appendix~\ref{app:washout}.

\section{Asymmetric Dark matter from $\eta$ decay}\label{Dark matter}
At temperature above its mass scale, vector-like fermion doublet $\psi$ remains in thermal equilibrium in the early Universe. DM candidate $\chi$, in spite of being a SM gauge singlet, attains thermal equilibrium via the process $\psi \psi \to \chi \chi$ shown in Fig.~\ref{sidm_ann} mediated by the light vector boson $Z'$. This is because the dark gauge coupling $g'$ is sufficiently large {($g' \sim 0.1$) as required for sufficient self-interaction (discussed in Sec.~\ref{sidm}). 

\begin{figure} [h!]
\centering
\includegraphics[width=0.3\textwidth]{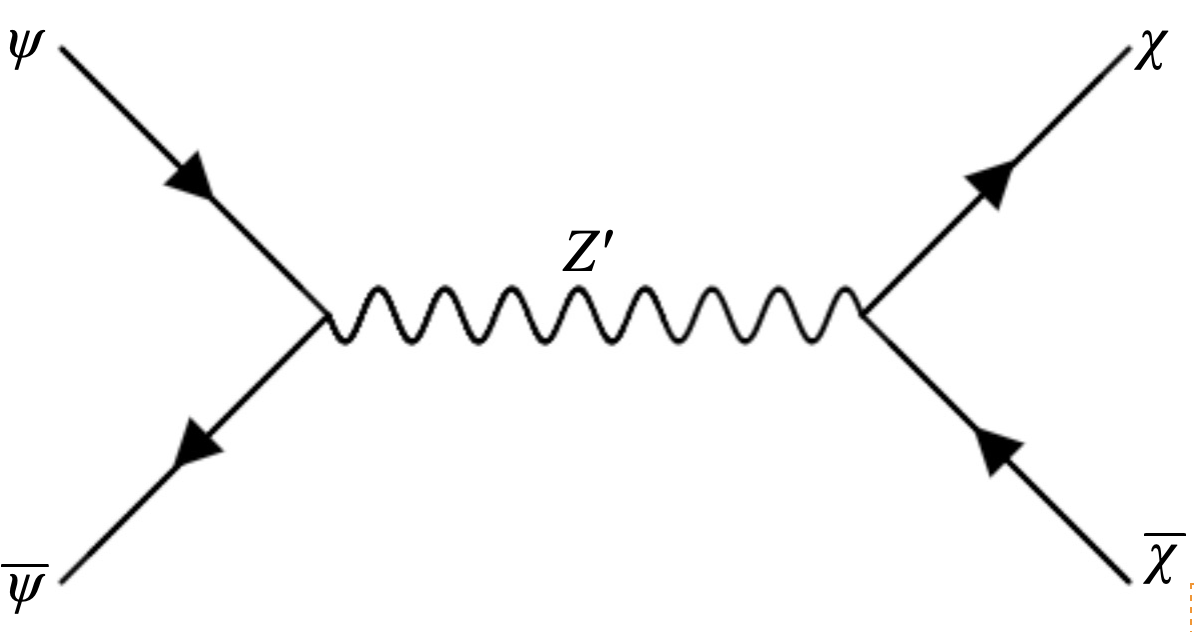}
\caption{Feynman diagram responsible for bringing DM into thermal equilibrium.}
\label{sidm_ann}
\end{figure}

However, the same gauge coupling also facilitates efficient DM $\chi$ annihilation into vector bosons $Z'$ through the process shown in Fig.~\ref{sidm_ann2}. 

\begin{figure} [h!]
\centering
\includegraphics[width=0.3\textwidth]{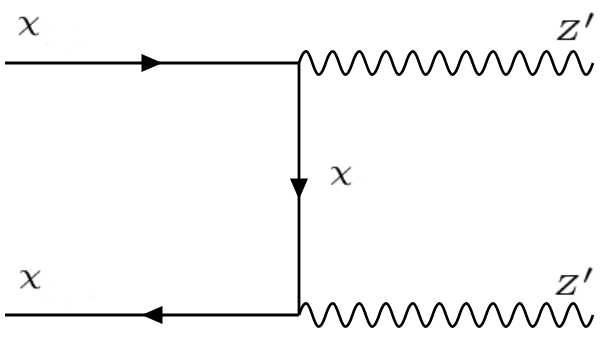}
\caption{Annihilation of the symmetric component of the DM.}
\label{sidm_ann2}
\end{figure}
Consequently, the relic density of the symmetric part becomes vanishingly under-abundant. Symmetric relic density can be quantified from the Boltzmann equation in terms of $Y^{sym}_{\chi}$ and $x=M_{\chi}/T$, as follows,
		\begin{equation}
		\label{b_sn}
\frac{dY^{sym}_{\chi}}{dx}= -\frac{s(M_{\chi})}{x^2  H(M_{\chi})} \langle\sigma v\rangle_{\overline{\chi}\chi \to~ Z' Z'} \left(Y^2_{\chi} -\left(Y^{eq}_{\chi}\right)^2\right)\,,
\end{equation}
where $\langle\sigma v\rangle_{\overline{\chi}\chi \to~ Z' Z'}$ is the thermally averaged cross-section for $\overline{\chi}\chi \to Z' Z'$, which is,
	\begin{equation}
		\langle\sigma v\rangle_{\overline{\chi}\chi \to~ Z' Z'} \approx \frac{\pi \alpha'^2}{M^2_{\chi}} \left( 1-\frac{M^2_{Z'}}{M^2_{\chi}}\right)^2
\end{equation}
where $\alpha'=g'^2/4\pi $. For $g' \sim 0.1$ and DM mass $M_{\chi} \sim 1 ~{\rm GeV}$, this leads to a cross-section which is about two orders of magnitude larger than the required ballpark to obtain correct relic density. We show the under-abundant relic density  of the symmetric component in Fig.~\ref{sidm_th}. 
\begin{figure}[ht!]
\includegraphics[width=90mm]{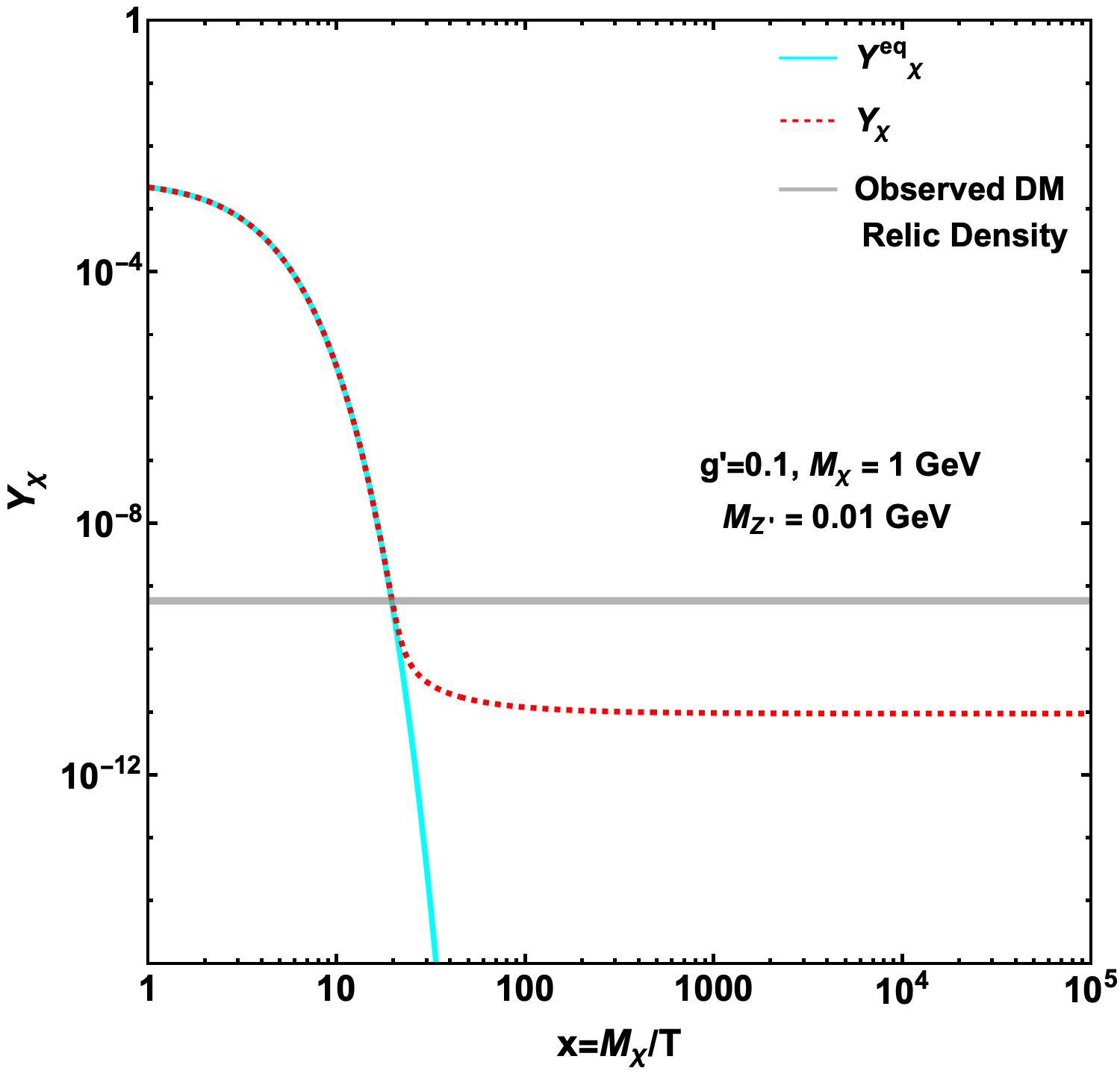}~
	  \caption{Under-abundant relic of symmetric component of the DM depicted by dotted red curve. The equilibrium number density is depicted by the cyan curve. Straight grey line shows the observed DM relic density.}
     \label{sidm_th}             
\end{figure}

CP-violating out of equilibrium decays $\xi_1^- \to \psi^- \chi, \,\xi_1^+ \to (\psi^-)^c \chi^c $ produce an asymmetry between $\chi$ and $\bar{\chi}$ as well as between $\psi$ and $\bar{\psi}$, similar to lepton asymmetry. The symmetric component of $\psi$ gets diluted by three annihilation processes: $\overline{\psi}\psi \to {\rm SM~SM}, \,\overline{\psi}\psi \to \overline{\chi}\chi,\, \overline{\psi}\psi \to Z' Z'$. Note that, in presence of only the process $\overline{\psi}\psi \to {\rm SM~SM}$ {\it i.e.,} via gauge couplings of $\psi$ to SM, $\psi$ remains under-abundant upto TeV scale, see Appendix~\ref{appendix1}. If we assume that $\psi\psi Z'$ coupling is of the same order as that of $\chi\chi Z'$, the other two channels dominates over $\overline{\psi}\psi \to {\rm SM~SM}$, and the symmetric doublet component remains many orders below the correct relic ballpark for GeV scale DM. The symmetric part of DM $\chi$ annihilates into pairs of $Z'$ as discussed above. Therefore, we can safely assume $\chi$ relic to be asymmetric, solely produced from CP-violating out of equilibrium decays $\xi_1^- \to \psi^- \chi, \,\xi_1^+ \to (\psi^-)^c \chi^c$. Feynman diagrams responsible for generation of asymmetries in dark sector are shown in Fig. \ref{DA}. The asymmetric number density of $\psi$ gets converted to a net $\chi$ density through the decay process: $\psi \to \chi \bar{f} f$ (we assume $M_\psi>M_\chi$), induced via the soft $\mathcal{Z}_2$ symmetry breaking term $\mu^2 H^\dagger \eta$. This decay does not wash out the dark 
asymmetry, as it preserves the dark charge $D(\psi)=D(\chi)$ and 
therefore acts only as a transfer process that converts the primordial 
$\psi$ asymmetry into the stable $\chi$.  The total dark asymmetry 
$Y_{\Delta D}=Y_{\Delta\psi}+Y_{\Delta\chi}$ is thus conserved exactly.  
We provide a detailed wash-out analysis considering all relevant interactions in Appendix~\ref{app:washout}.

\begin{figure} [h!]
\centering
\includegraphics[width=90mm]{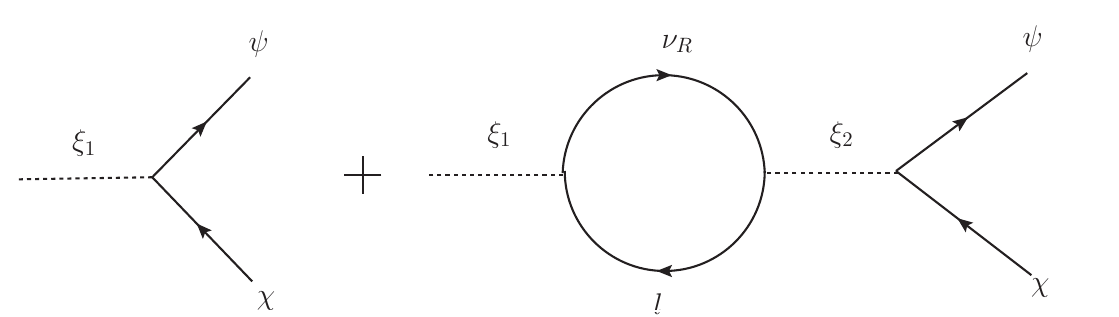}
\caption{Tree level and self energy correction diagrams producing the dark matter asymmetry.}
\label{DA}
\end{figure}
The amount of CP-asymmetry can be quantified as,
\begin{eqnarray}
\epsilon_\chi &=& [Br(\xi_{1}^{-} \rightarrow \psi^{-} \chi)-Br(\xi_{1}^{+}\rightarrow (\psi^{-})^{c} \chi^{c})]\nonumber\\
    &=& \frac{Im\left(\lambda_{1}^{*} \lambda_{2} \mathop{\sum_{k,l}} f^{*}_{1kl} f_{2kl} \right)}{8\pi^2(M_{2}^{2}-M_{1}^{2})} \left[\frac{M_{1}^{2}M_{2}}{\Gamma_{1}} \right]=-\epsilon_L.
 \label{DA_asy2}
 \end{eqnarray}
 
 \begin{figure}[h!]
				\includegraphics[width=90mm]{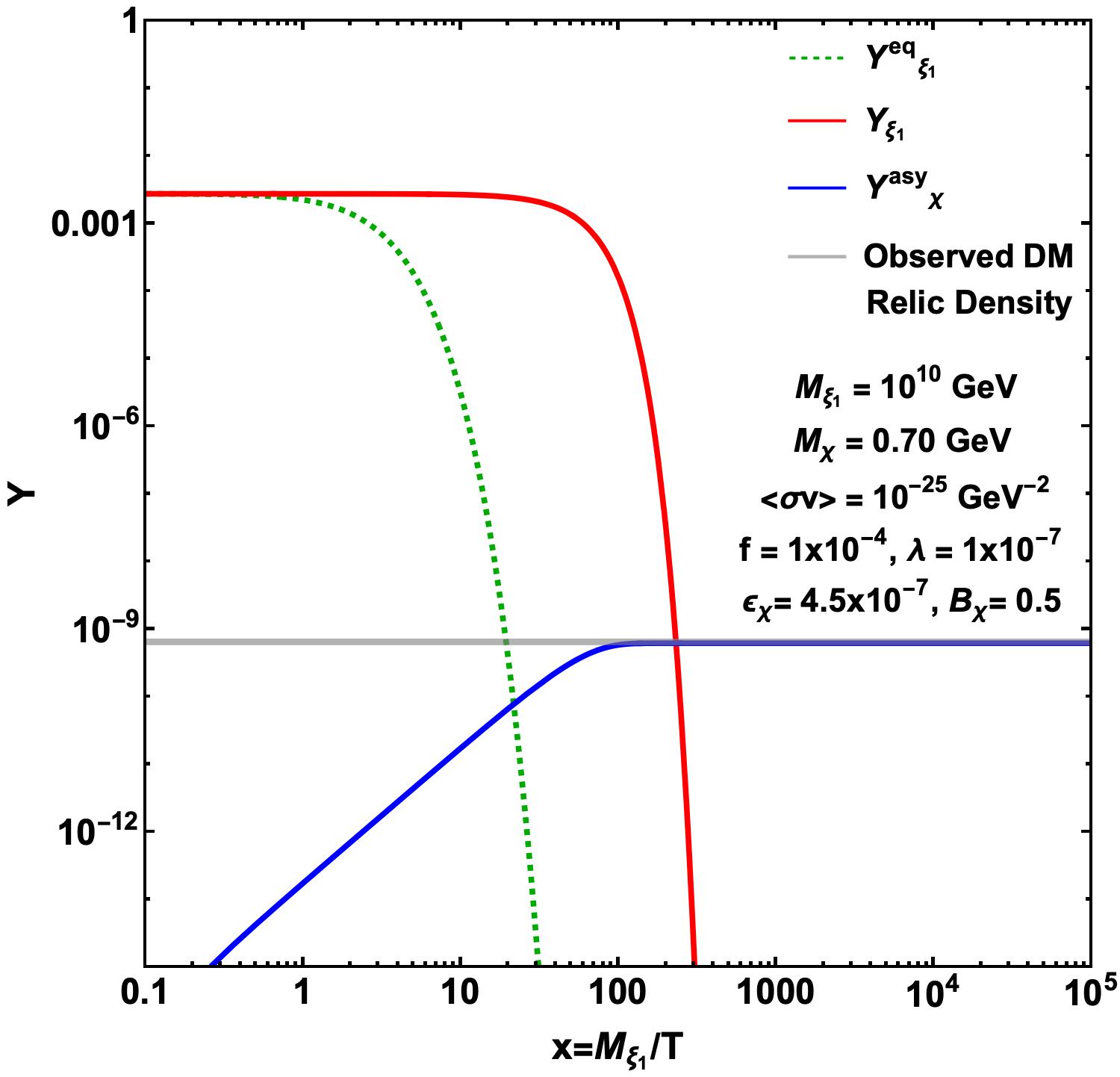}~
				 \caption{Relic density of asymmetric component of the DM obtained from the decay of $\eta$.}
                   \label{relic_asy}             
\end{figure}

The abundance of asymmetric $\chi$ density can be estimated from the coupled Boltzmann equations: 
\begin{equation}
\label{eqn.1}
\begin{split}
\frac{dY_{\xi_{1}}}{dx}&=-\frac{x}{H(M_{\xi_{1}})}s<\sigma|v|_{(\xi_{1}\xi_{1} \rightarrow All)}>\left[ Y_{\xi_{1}}^{2}-{Y_{\xi_{1}}^{eq}}^2 \right]\\
&-\frac{x}{H(M_{\xi_{1}})}\Gamma_{(\xi_{1} \rightarrow All)} \left[ Y_{\xi_{1}}-Y_{\xi_{1}}^{eq} \right]\,,\\
\frac{dY_{\chi-\overline{\chi}}}{dx}&=\epsilon_\chi \frac{x}{H(M_{\xi_{1}})}\Gamma_{(\xi_{1}\rightarrow All)} B_{\chi} \left[ Y_{\xi_{1}}-Y_{\xi_{1}}^{eq} \right].
\end{split}
\end{equation}
Here we use $|\epsilon_L|=|\epsilon_\chi|= 4.5 \times 10^{-7}$. Other parameters are also kept same as that in Sec.~\ref{Dirac neutrino mass} and Sec.~\ref{Production of LA} {\it i.e.,} $f = 10^{-4}, \,\lambda = 10^{-7},\, \sigma|v|_{(\xi_{1}\xi_{1} \rightarrow All)}=10^{-25}\, {\rm GeV}^{-2}$. With this, we obtain the observed relic density for $M_\chi = 0.70~{\rm GeV}$ as shown in Fig.~\ref{relic_asy}. 
The solid red curve shows the evolution of the number density of $\xi_{1}$ particles and the solid blue line depicts the evolution of $Y^{\rm asy}_{\chi}$. Note that the observed baryon asymmetry does not get affected by the decay $\xi_1 \to \psi \chi$, since $\psi$ and $\chi$ being vector-like fermions, do not get converted to baryon asymmetry by the sphalerons~\cite{Arina:2011cu,Arina:2012fb,Arina:2012aj,Narendra:2017uxl}. 


\section{Dark Matter Self-interaction}\label{sidm}
	To alleviate the small-scale anomalies of $\Lambda{\rm CDM}$, the typical DM self-scattering cross-section should be $\sigma \sim 1~{\rm cm}^2 /{\rm g} \approx 2\times 10^{-24}~{\rm cm}^2 /{\rm GeV}$, which is 14 orders of magnitude larger than the typical WIMP cross-section($\sigma \sim 10^{-38} \, {\rm cm}^2/GeV$). Besides, the self-interaction is desired to be stronger for smaller DM velocities such that it can have a large impact on small scale structures. On the other hand, it must get reduced for larger DM velocities to remain consistent with large scale $\Lambda{\rm CDM}$ predictions. SIDM with an MeV scale mediator fulfils all this requirements. $U(1)_D$ vector boson $Z'$ in our model serves this purpose. The term $g' \overline{\chi}\chi Z'$ given in Eq.\,\ref{Eq-Lagrangian} enables DM self-interaction. Self-interaction can be represented by the Fynman diagram in Fig.~\ref{feyn_self}.
\begin{figure} [h!]
\centering
\includegraphics[width=0.25\textwidth]{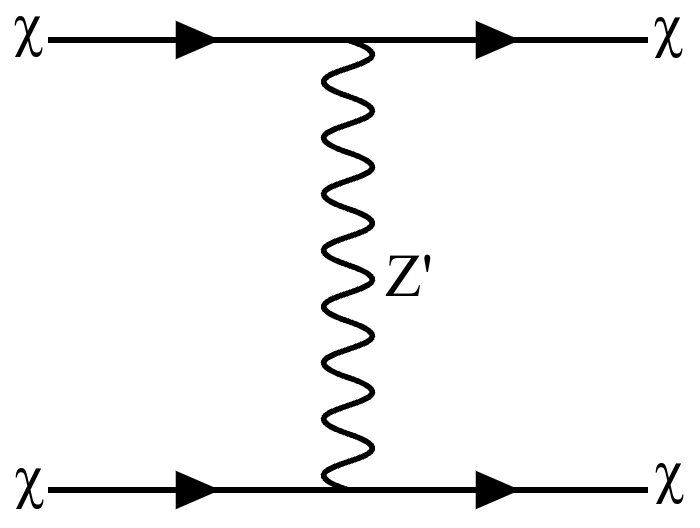}
\caption{Feynman diagram for DM self-interaction.}
\label{feyn_self}
\end{figure}

In non-relativistic limit, scattering can be described by Yukawa-type potential,
	\begin{equation}
		V(r)= \pm \frac{g'^2}{4\pi r}e^{-M_{Z'} r}
	\end{equation} 
	where, the + (-) sign denotes repulsive (attractive) potential. For vector mediators, $\chi \chi$ scattering is repulsive, which is the relevant case for this work. To capture the forward scattering divergence, we define transfer cross-section $\sigma_T$ as~\cite{Feng:2009hw,Tulin:2013teo,Tulin:2017ara},
	\begin{equation}
		\sigma_T = \int d\Omega (1-\cos\theta) \frac{d\sigma}{d\Omega}
	\end{equation}
Three distinct regimes can be identified for the scattering.
\begin{itemize}
\item Born regime $\left(g'^2 M_\chi/(4\pi M_{Z'}) \ll 1, \frac{M_\chi v}{M_{Z'}}\geq 1\right)$
\item Classical regime $\left(g'^2 M_\chi/(4\pi M_{Z'})\geq 1, \frac{M_\chi v}{M_{Z'}}\geq 1\right)$
\item Resonance Regime $\left(g'^2 M_\chi/(4\pi M_{Z'})\geq 1, \frac{M_\chi v}{M_{Z'}}\leq 1\right)$
\end{itemize}
Cross-sections for each regime are listed in Appendix~\ref{appendix2}. Perturbative calculations hold good only in the Born regime. In the resonant regime, for attractive potential we see quantum mechanical resonances for smaller velocities due to (quasi)-bound state formation~\cite{Dutta:2022knf}. Since in the present case, we have asymmetric DM with an repulsive potential, we do not see any resonance spikes. In Fig.~\ref{sidm1}, we show the self-interaction allowed parameter space for $g'=0.1$ in $M_\chi$ - $M_{Z'}$ plane. We constrain $\sigma/M_\chi$ as:
\begin{itemize}
\item $0.1-10~{\rm cm}^2/{\rm g}$ for dwarfs ($v\sim 10~ \rm km/s$) 
\item $0.1-1~{\rm cm}^2/{\rm g}$ for galaxies ($v\sim 100~ \rm km/s$) 
\item $0.1-1~{\rm cm}^2/{\rm g}$ for clusters ($v\sim 1000~ \rm km/s$)  
\end{itemize}
These regions are depicted by cyan, green and magenta colored regions in Fig.~\ref{sidm1} as indicated in figure inset.
	\begin{figure}[h!]
	\centering
		\includegraphics[width=9cm]{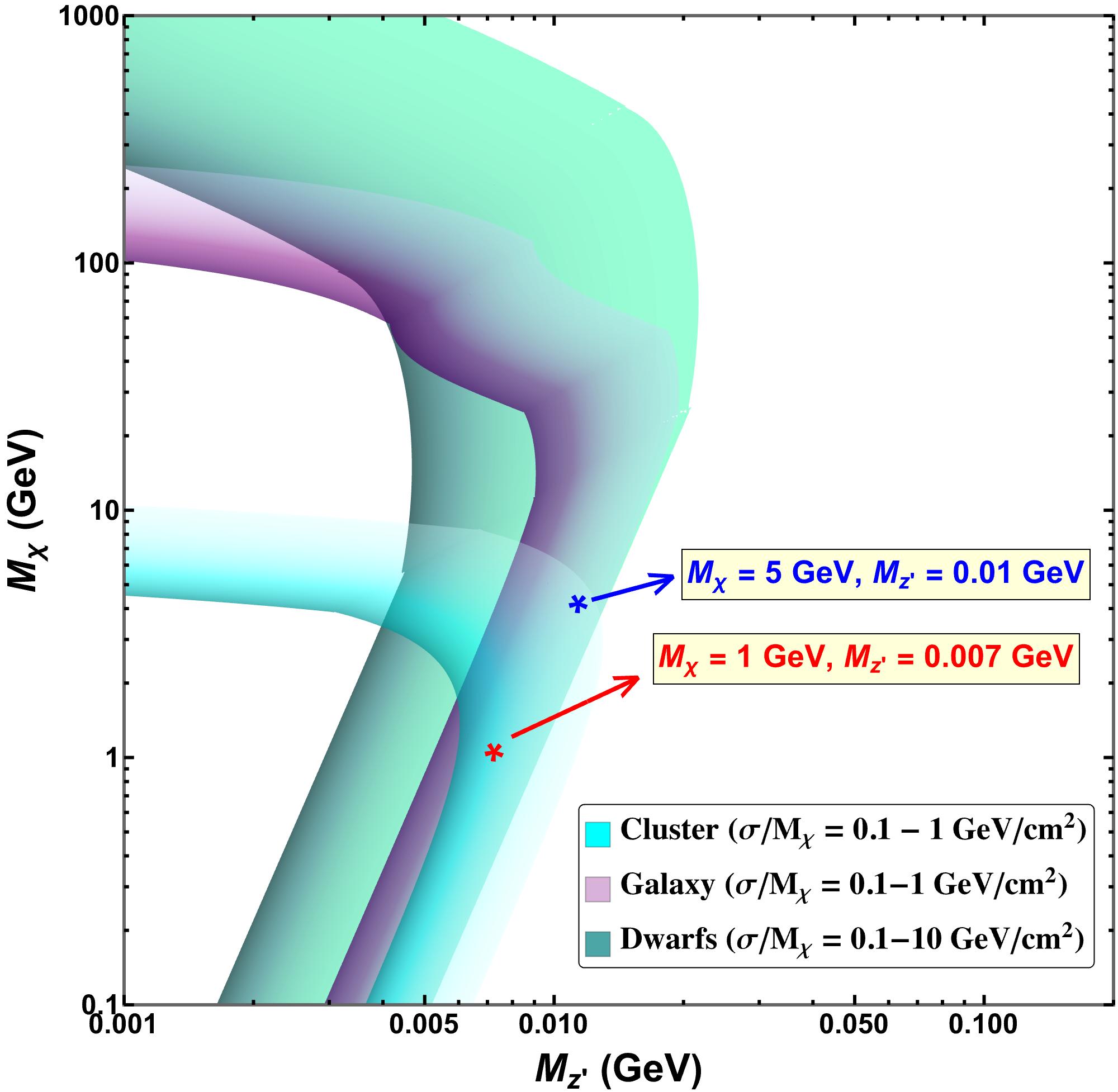}
		\caption{Parameter space for self-interaction in $M_\chi$ - $M_{Z'}$ plane with $g'=0.1$.}
		\label{sidm1}
	\end{figure}	
\begin{figure}[h!]
	\centering
			\includegraphics[width=9cm,height=7.5cm]{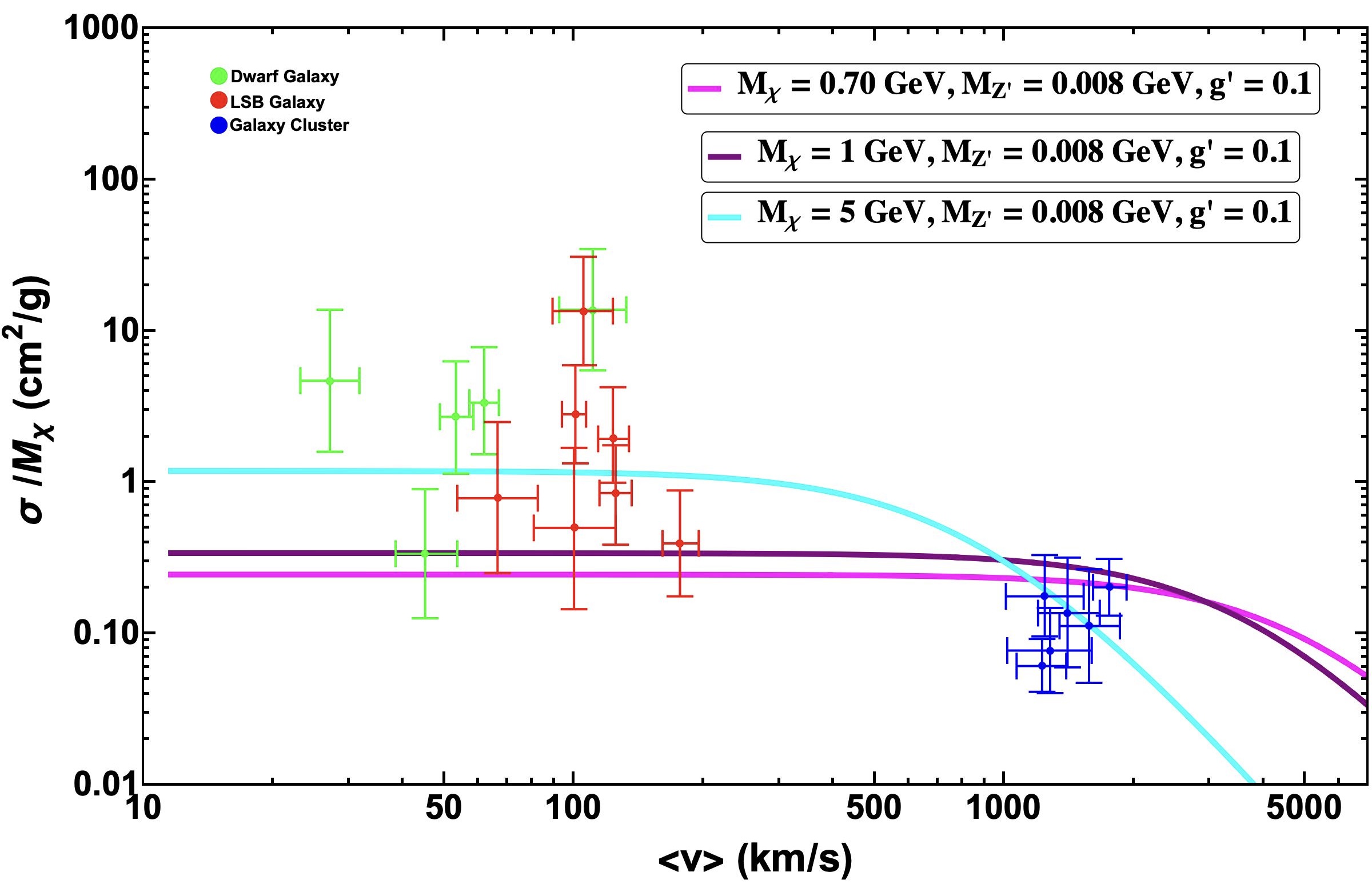}
		\caption{Self-interaction cross-section per unit mass of DM in terms of average collision velocity.}
		\label{sidm2}
	\end{figure}	
		
	The top corner corresponds to the Classical regime while the bottom corner corresponds to the Born regime, with resonant regime sandwiched between the two regimes. 
	
	 We also plot self-scattering cross-section per unit DM mass versus average collision velocity in Fig.~\ref{sidm2} for different DM masses. Keeping $M_{Z'}= 0.008~{\rm GeV}$ and $g'=0.1$, we plot the cross-section per unit DM mass for $M_\chi=0.7~{\rm GeV}, 1~{\rm GeV}, 5~{\rm GeV}$. It fits data from dwarfs (red), low surface brightness (LSB) galaxies (blue), and clusters (green)~\cite{Kaplinghat:2015aga, Kamada:2020buc}, for the values of $M_\chi$, $M_{Z'}$ as indicated in the figure. It is clear that the model can appreciably explain velocity-dependent DM self-interaction.

\section{DM Direct Detection} \label{sidm_dd}
DM $\chi$ undergoes elastic scattering off detector nucleons. This is facilitated by $Z-Z'$ mixing as shown in Fig.~\ref{direct}.

\begin{figure}[h!]
\centering
\includegraphics[scale=0.20]{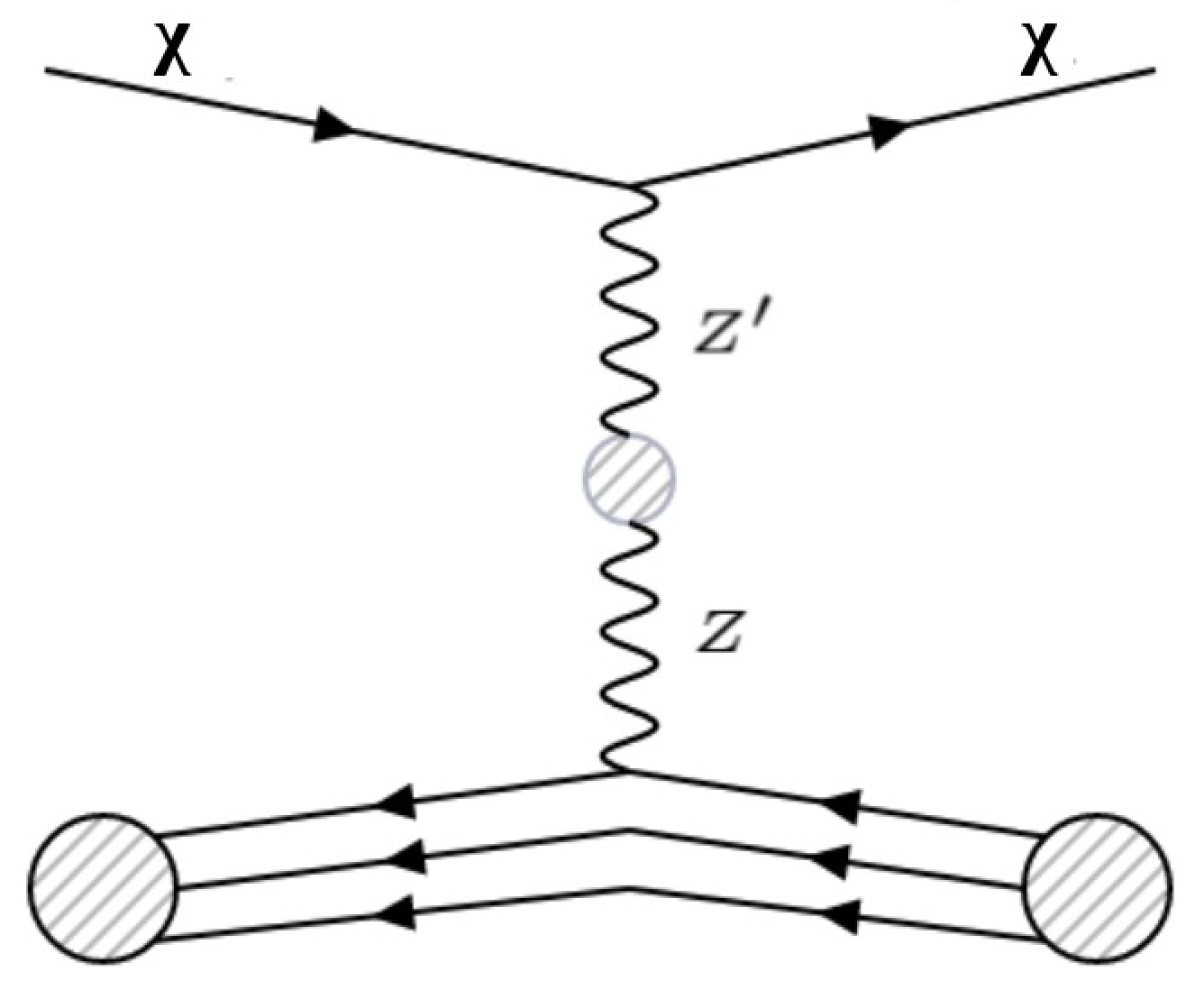}
\caption{Elastic scattering of DM $\chi$ off a nucleon mediated by $Z-Z'$ mixing.}
\label{direct}
\end{figure}


\begin{figure}[h!]
\centering
\includegraphics[width=90mm]{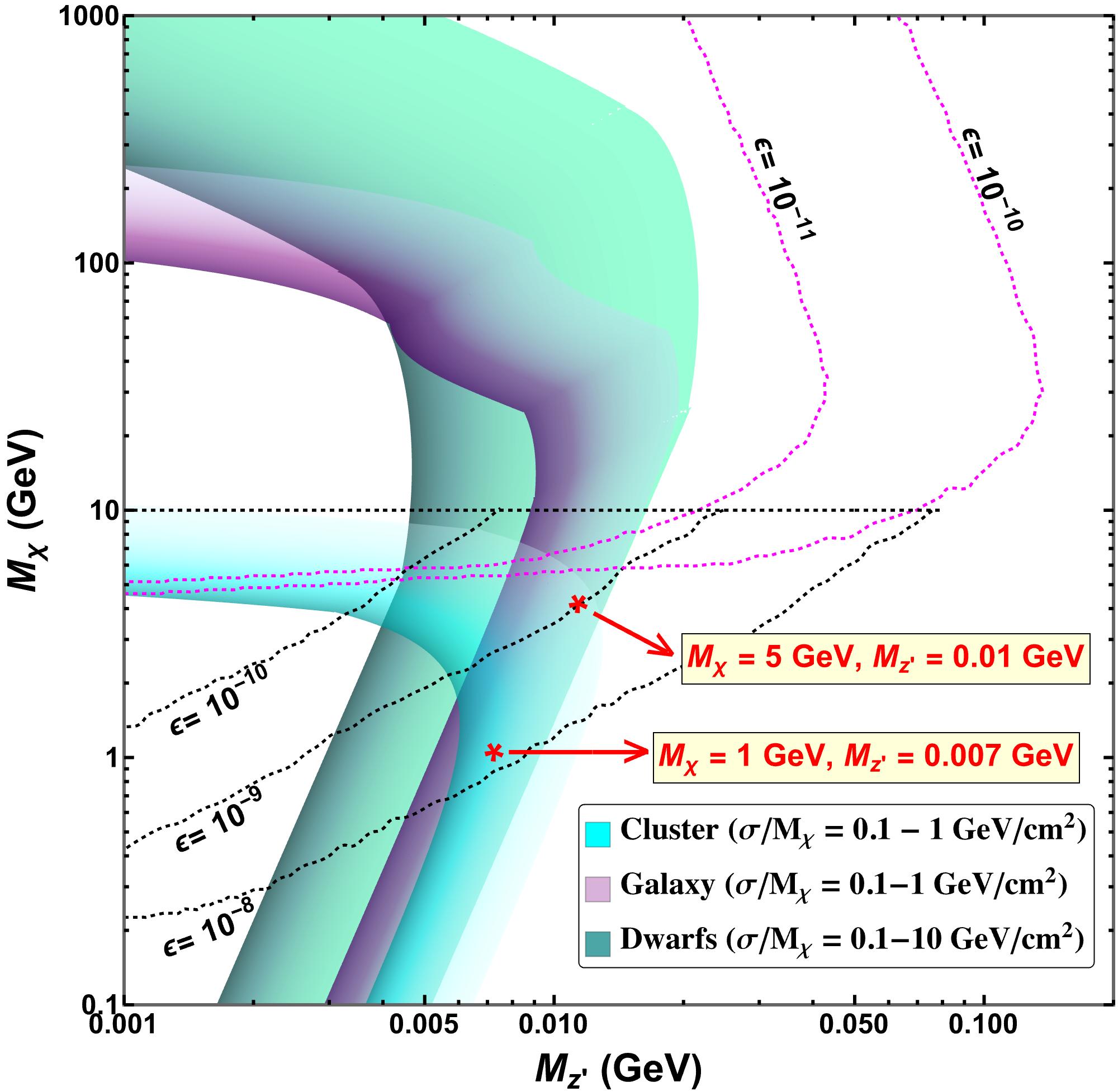}
\caption{Constraints from DM direct detection in the plane of DM mass $(M_{\chi})$ versus mediator mass $(M_{Z'})$ with $g'=0.1$.}
\label{sidmdd}
\end{figure}

Let us consider spin independent direct search, the cross-section for which is given by,
\begin{equation}
\sigma^{\rm SI}_{\chi N}=\frac{g^2 g'^2 \epsilon^2}{\pi}\frac{ \mu^2_{\chi N}}{M^4_{Z'}} \frac{\big( Z f_p + (A-Z) f_n \big)^2}{A^2}.
\end{equation}
where, $\mu_{\chi N} = \frac{M_\chi m_N}{(M_\chi+m_N)}$ is the reduced mass of the DM-nucleus ($N$) system with proton number $Z$ and mass number $A$. $\epsilon$ is the $Z-Z'$ kinetic mixing, $f_p$ ($f_n$) are interaction strengths for proton (neutron) respectively.

We constrain model parameters with data from direct search experiments like CRESST-III \cite{Abdelhameed:2019hmk} and XENON1T \cite{Aprile:2018dbl}. In Fig.~\ref{sidmdd}, the most stringent constraints from CRESST-III (dotted contours in black) and XENON1T (dotted contours in pink) experiments are shown assuming $g'=0.1$. For below (above) $10 {\rm GeV}$ DM, we use constraints from CRESST-III (XENON1T) experiment. The region towards left of each contour is excluded for the mixing parameter mentioned alongside the curves.

\section{Collider Phenomenology}
\label{coll}

In this section we summarise the collider phenomenology of the dark
fermion $\Psi$ and the dark gauge boson $Z'$.

\vspace{0.5cm}
\noindent {\bf Absence of displaced vertices:}
\vspace{0.25cm}

In the low-energy effective theory, the three-body decay
$\psi \rightarrow \chi f\bar f$ is induced by integrating out the heavy
scalar doublet $\xi_i$, generating the dimension-6 operator
\begin{equation}
\mathcal{L}_{\rm eff} \supset 
\frac{y_1 y_2}{M_{\xi_i}^2}
(\bar\chi\,\Gamma\,\psi)(\bar f\,\Gamma\,f),
\end{equation}
where $f$ is a Standard Model fermion.  Defining
\begin{equation}
y_{\rm eff}^2 \equiv y_1 y_2,
\qquad
\Lambda \equiv M_{\xi_i},
\end{equation}
the decay width is
\begin{equation}
\Gamma_\psi \simeq 
\frac{y_{\rm eff}^4}{192\pi^3}
\frac{M_\psi^5}{\Lambda^4},
\label{eq:widthPsiCollider}
\end{equation}
The proper decay length ($L_0=c\tau_\psi=\hbar c/\Gamma_\psi$) is obtained as,
\[
L_0 \simeq
1.17\times 10^{13}\ \text{m}\;
\left(\frac{1}{y_{\rm eff}}\right)^{4}
\left(\frac{\Lambda}{10^{10}\ \text{GeV}}\right)^{4}
\left(\frac{1\ \text{TeV}}{M_\psi}\right)^{5},
\]
so that even for $y_{\rm eff}=1$, $M_\psi=1~\mathrm{TeV}$ and 
$\Lambda=10^{10}~\mathrm{GeV}$ we find 
$L_0 \simeq 1.17\times 10^{13}~\mathrm{m}$,
far exceeding the size of any collider detector. For smaller $y_{\rm eff}$ or smaller $\psi$ mass, the lifetime (and hence the decay length) increases
further. Thus throughout the entire region relevant for successful
visible and dark asymmetry generation, $\psi$ is effectively stable at
colliders; no displaced vertices are expected. The collider signature
of $\psi$ is, therefore, that of a heavy long-lived particle. The charged component $\psi^-$ behaves as a heavy stable
charged particle (HSCP), producing a slow, highly ionising track
through the tracker and muon system, while the neutral component $\psi^0$ behaves as missing transverse
energy (MET). The production channels, $
pp \to \psi^+ \psi^- , 
pp \to \psi^\pm \psi^0 , 
pp \to \psi^0 \psi^0$ thus lead to a characteristic mixture of HSCP tracks and MET, a hallmark
signature of collider-stable electroweak doublets. Current CMS and ATLAS HSCP searches~\cite{CMS:2010HSCP,ATLAS:2019HSCP} 
exclude long-lived charged fermions up to $1.3\text{--}1.6~\mathrm{TeV}$. Neutral components contribute to MET+ISR signatures, constrained by
monojet searches~\cite{ATLAS:2021monojet,CMS:2021monojet},  
which give a weaker bound $
M_\psi \gtrsim 100\text{--}200~\mathrm{GeV}.
$

The dark gauge boson $Z'$ is light ($M_{Z'}\sim\mathrm{MeV}$) and the
dark coupling is sizeable ($g'\sim0.1$), as required for
self-interacting dark matter.  However, the interactions
of $Z'$ with the Standard Model are controlled by the kinetic
mixing, which is extremely small in the phenomenologically relevant
region (see Fig.~14):
\[
\epsilon \lesssim 10^{-8},
\qquad 
g'_{\rm eff} = \epsilon\,g' \sim 10^{-9}
\]
Such a light and ultra–feebly coupled vector is far below the sensitivity 
of all current LHC dilepton, monophoton, dijet, or invisible-$Z$ searches
\cite{ATLAS:2019llZ,CMS:2021monojet,CMS:2021monophoton}.
Furthermore, as demonstrated in Appendix~\ref{app:washout}, all portal-induced
scatterings between dark and visible fermions satisfy
\begin{equation}
\frac{\Gamma_{\rm scatt}}{H}
\sim 10^{-20}
\qquad \text{at } T = M_\Psi,
\end{equation}
ensuring that the two sectors never thermalise after the epoch of
asymmetry generation. 
This not only guarantees the preservation of the asymmetry, but also explains why the $Z'$ does not produce
observable collider signatures. Hence, in this set-up, displaced vertices and successful
asymmetry generation are mutually incompatible, which is a distinctive feature of the model.

\section{Conclusions}\label{Conclusion}
As no consensus has been established regarding Dirac or Majorana nature for neutrinos yet, we, in this work, explain baryon and dark matter asymmetry in a common framework assuming assume neutrinos to be Dirac particles. 
We extend SM with a $SU(2)_L$ doublet $\psi$ and a singlet $\chi$. The singlet fermion $\chi$ being odd under the discrete $\mathcal{Z}_2$ symmetry behave as a candidate of DM. The same $\mathcal{Z}_2$ symmetry disallowed neutrino Dirac mass. $\mathcal{Z}_2$ symmetry is softly broken without destabilizing DM $\chi$, ensured by $M_\chi < M_\psi$. This also leads to a Dirac mass of neutrinos protected by a global $B-L$ symmetry. A heavy scalar doublets ($\eta$), having non-trivial $\mathcal{Z}_2$ charge, undergoes out-of-equilibrium decay. Decay channels $\eta \to \nu_R\ell$ and $\eta\to \chi \psi$ generates lepton and dark matter asymmetries. As $B-L$ is an exact symmetry, the CP-violating decay of $\eta$ to $ \nu_R\ell$ produces equal and opposite $B-L$ asymmetries in the left and right-handed sectors. The right-handed sector is weakly coupled to SM sector, required by neutrino Dirac mass. $B-L$ asymmetry in the left-handed sector converts to a net $B$-asymmetry via the $B+L$ violating sphaleron transitions. The asymmetries in the right-handed sector remains intact. The asymmetries neutralizes much after the electroweak phase transition when the sphaleron transitions gets suppressed. Meanwhile the symmetric DM component annihilates away to pairs of light mediators. The light mediators facilitates self-interaction among the DM particles, a essential feature for alleviating the small-scale $\Lambda{\rm CDM}$ anomalies. The kinetic mixing between dark and visible sectors facilitates direct detection of DM. We have found the relevant parameter space for SIDM confronted with direct search constraints from experiments like CRESST-III and XENON-1T. Since kinetic mixing suppresses all annihilations to SM final states, all gamma, cosmic, and neutrino fluxes are well below the near future reach of indirect probes~\cite{Arina:2018zcq,https://doi.org/10.1002/andp.19163561802,Cassel:2009wt,Iengo:2009ni,Slatyer:2009vg}. The indirect signals of SIDM remains much below the current constraints by the experiments like Fermi-LAT~\cite{Fermi-LAT:2013sme,Fermi-LAT:2015att}, MAGIC~\cite{MAGIC:2016xys}, HESS~\cite{HESS:2018cbt}, AMS-02~\cite{PhysRevLett.110.141102}, CMB by Planck~\cite{Aghanim:2018eyx}, and $ \gamma $-rays by INTEGRAL~\cite{Knodlseder:2007kh}.

Although direct production of either $\psi$ or $Z'$ at high-energy 
colliders is highly suppressed, the model exhibits several distinctive 
phenomenological features.  
Both the visible and dark asymmetries originate from the same CP-violating 
decays of the heavy scalar doublets, implying correlated phases and a 
common dynamical origin of baryogenesis and dark-genesis.  
The dark sector contains a self-interacting MeV-scale vector mediator $Z'$ 
with an ultra-feeble kinetic portal to the Standard Model, while the heavy 
dark fermion $\psi$ is effectively stable on detector scales.
This characteristic combination of {\it `self-interacting dark matter + ultra-feeble
portal + collider-stable $\psi$'} is not realised in typical ADM or leptogenesis scenarios and therefore provides a distinguishing phenomenological
signature of the framework. At future high-energy colliders such as the ILC~\cite{ILC:2013}, 
CLIC~\cite{CLIC:2018}, FCC-hh~\cite{FCC:2019}, CEPC~\cite{CEPC:2018}, 
and SPPC~\cite{SPPC:2015}, the pair-production of $\psi$ can be significantly enhanced.  
the charged component of 
$\psi$ would appear as a heavy stable charged particle (HSCP), whereas the 
neutral component would manifest as a large missing-energy signal.  
Although $Z'$ remains inaccessible due to its tiny kinetic mixing,
the observation of long-lived $\psi$ states, together with astrophysical 
measurements of dark self-interactions would provide complementary avenues 
to probe the underlying dark sector and the unified origin of visible and 
dark asymmetries in this model.
  



\appendix

\section{Relic Abundance of Doublet DM having only gauge interactions with the SM}
\label{appendix1}
If the DM is a vector-like fermion doublet $\psi$ of the form $\psi^T=(\psi^0, \psi^-)$ (with hypercharge 
$Y=-1$, where we use $Q=T_3+Y/2$), odd under an imposed $\mathcal{Z}_2$ symmetry, its neutral component becomes stable DM.
The Lagrangian for the doublet DM is given by,
\begin{equation}
\label{doublet_Lagrangian}
\mathcal{L} = \overline{\psi} \left( i\gamma^\mu D_\mu - M_\psi \right) \psi 
\end{equation}   
where, the covariant derivative involving the $SU(2)_L$ gauge boson triplet $W_\mu^a$ ~$(a=1,2,3)$ and $U(1)_Y$ gauge boson $B_\mu$ given by:
 \begin{equation}
 D_\mu = \partial_\mu - i\frac{g}{2}\tau_a.W_\mu^a - ig_Y\frac{Y}{2}B_\mu
 \end{equation}
where $\tau_a$ are Pauli spin matrices (generators of $SU(2)$), $g$ and $g_Y$ denote $SU(2)$ and $U(1)$ coupling strengths respectively.

\begin{figure}[h!]
	\centering
	\includegraphics[height=7cm,width=7cm]{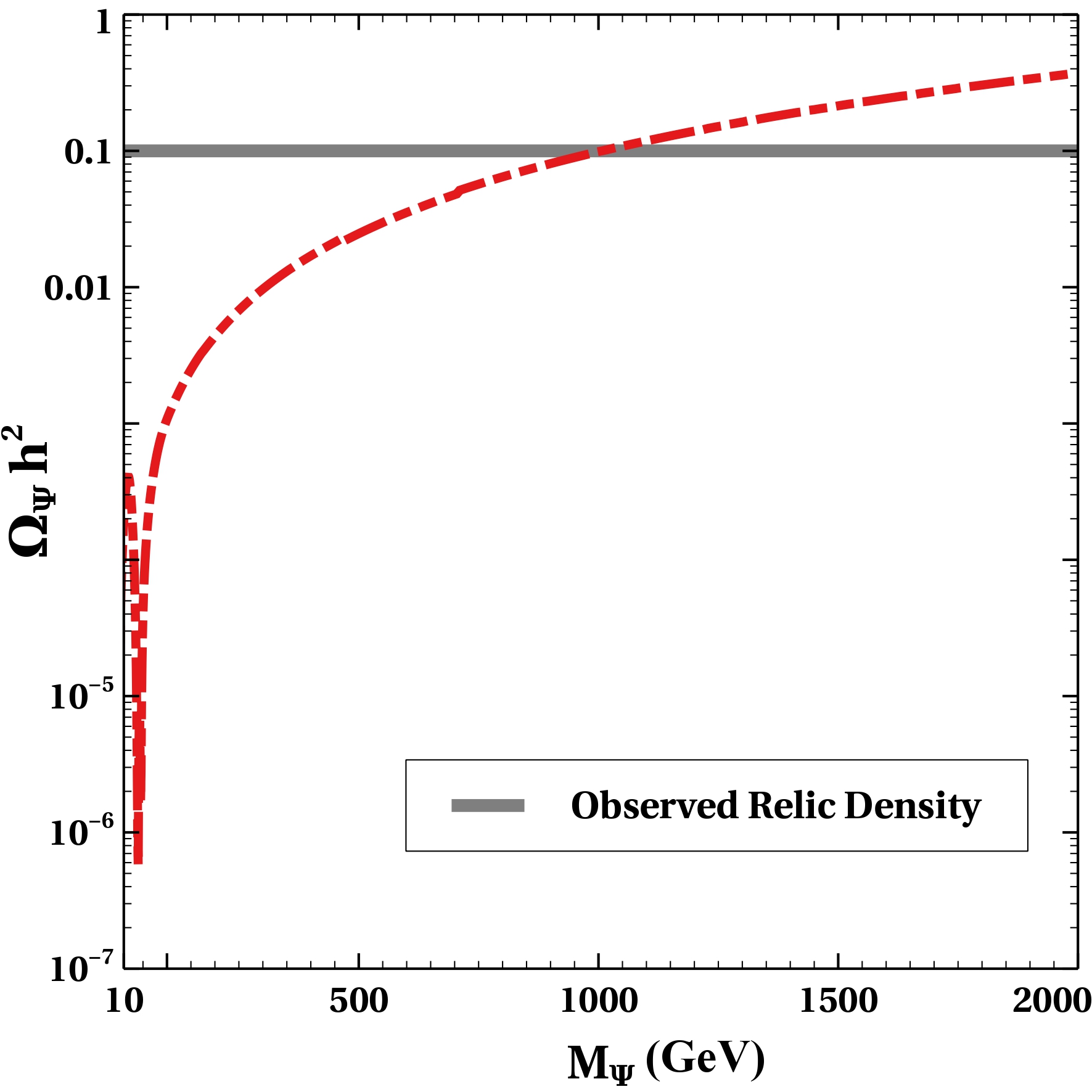}
	\caption{Relic density of doublet DM with only gauge interaction with SM.}
	\label{doublet_relic}
\end{figure}

Expanding the covariant derivative of the Lagrangian given in Eq.\ref{doublet_Lagrangian}, we get the interaction term of $\psi^0$ and $\psi^\pm$ with the SM gauge bosons as follows:
\begin{equation}
\label{doublet_lag}
\begin{aligned}
 \mathcal{L}_{int} &=\overline{\psi}i\gamma^\mu(-i\frac{g}{2}\tau.W_\mu - ig_Y\frac{Y}{2}B_\mu )\psi
 \\ 
   &= \Big(\frac{e}{2\sin\theta_W \cos\theta_W}\Big)\overline{\psi^0}\gamma^\mu Z_\mu \psi^0
   \\
   & +\frac{e}{\sqrt{2}\sin\theta_W}(\overline{\psi^0}\gamma^\mu W^+_\mu \psi^- + \psi^+\gamma^\mu W^-_\mu \psi^0) \\
   & - e \psi^+\gamma^\mu A_\mu \psi^-\\ 
    & -\Big(\frac{e \cos2\theta_W}{2\sin\theta_W \cos\theta_W}\Big)~ \psi^+\gamma^\mu Z_\mu \psi^- .
\end{aligned}
\end{equation}
where $g = \frac{e}{\sin\theta_W}$ and $g_Y= \frac{e}{\cos\theta_W}$, $e$ and $\theta_W$ are the elementary charge and Weinberg angle respectively.

All interactions being fixed by the SM gauge charge, the model does not have any free parameter except the mass $M_\psi$. So both relic density and direct search cross-sections become a function of DM mass alone. The model has been implemented in \texttt{LanHEP}~\cite{SEMENOV1998124} and \texttt{MicrOmegas}~\cite{Belanger:2013oya}. The relic density obtained from the model as a function of DM mass is shown in Fig.~\ref{doublet_relic}. We observe that correct relic density is obtained only for DM mass around 1 TeV.

\section{DM Self-interaction Cross-sections at Low Energy}
	\label{appendix2}
	In the Born Limit ($\alpha' M_\chi/M_{Z'}<< 1$), for both attractive as well as repulsive potentials, the transfer cross-section is
		\begin{equation}
			\sigma^{\rm Born}_T = \frac{8 \pi \alpha'^2}{M^2_\chi v^4} \Bigg(\ln(1+ M^2_\chi v^2/M^2_{Z'} ) - \frac{M^2_\chi v^2}{M^2_{Z'}+ M^2_\chi v^2}\Bigg).
		\end{equation} 
		Outside the Born regime ($\alpha' M_\chi /M_{Z'} \geq 1 $), we have two distinct regions {\it viz} the classical region and the resonance region. In the classical limit ($M_\chi v/M_{Z'}\geq 1$), the solutions for an attractive potential is given by~\cite{Tulin:2013teo,Tulin:2012wi,Khrapak:2003kjw}
		\begin{equation}
			\sigma^{\rm classical}_T (\rm Attract.)=\left\{
			\begin{array}{l}
				
				\frac{4\pi}{M^2_{Z'}}\beta^2 \ln(1+\beta^{-1}) ;~~ \beta \leqslant 10^{-1}\\
				\frac{8\pi}{M^2_{Z'}}\beta^2/(1+1.5\beta^{1.65}) ;10^{-1}\leq \beta \leqslant 10^{3}\\
				\frac{\pi}{M^2_{Z'}}( \ln\beta + 1 -\frac{1}{2}ln^{-1}\beta); ~~\beta \geq 10^{3}\\
			\end{array}
			\right.
		\end{equation}  
		
In the classical limit ($M_\chi v/M_{Z'}\geq 1$), the solutions for repulsive potential is given by\cite{Tulin:2013teo,Tulin:2012wi,Khrapak:2003kjw}
		
		\begin{equation}
			\sigma^{\rm classical}_T (\rm Repul.)=\left\{
			\begin{array}{l}
				
				\frac{2\pi}{M^2_{Z'}}\beta^2 \ln(1+\beta^{-2}) ;~~\beta \leqslant 1\\
				\frac{\pi}{M^2_{Z'}}(\ln 2\beta^2 -\ln ~\ln 2\beta)^2 ;~~ \beta \geq 1\\
			\end{array}
			\right.
		\end{equation}   
		where $\beta = 2\alpha' M_{Z'}/(M_\chi v^2)$.

	In the resonant regime ($g'^2 M_\chi/(4\pi M_{Z'}) \geq 1, M_\chi v/M_{Z'}  \leq 1$), the quantum mechanical resonances and anti-resonance in $\sigma_T$ appear due to (quasi-)bound states formation in the attractive potential. In the resonant regime, non-perturbative results obtained by approximating the Yukawa potential to be a Hulthen potential $\Big( V(r) = \pm \frac{g'^2}{4\pi}\frac{ \delta e^{-\delta r}}{1-e^{-\delta r}}\Big)$ is given by~\cite{Tulin:2013teo}:	
	\begin{equation}
			\sigma^{\rm Hulthen}_T = \frac{16 \pi \sin^2\delta_0}{M^2_\chi v^2}
		\end{equation}
		
		where $l=0$ phase shift is given in terms of the $\Gamma$ functions by
		\begin{equation}
			\delta_0 ={\rm arg} \Bigg(\frac{i\Gamma \Big(\frac{i M_\chi v}{k M_{Z'}}\Big)}{\Gamma (\lambda_+)\Gamma (\lambda_-)}\Bigg)
			\end{equation} 
	where,		\begin{equation}
			\lambda_{\pm} = \left\{
			\begin{array}{l}
				1+ \frac{i M_\chi v}{2 k M_{Z'}}  \pm \sqrt{\frac{\alpha_D M_\chi}{k M_{Z'}} - \frac{ M^2_\chi v^2}{4 k^2 M^2_{Z'}}} ~~~~ {\rm Attractive}\\
				1+ \frac{i M_\chi v}{2 k M_{Z'}}  \pm i\sqrt{\frac{\alpha_D M_\chi}{k M_{Z'}} + \frac{ M^2_\chi v^2}{4 k^2 M^2_{Z'}}} ~~~~ {\rm Repulsive}\\
			\end{array}
			\right.
		\end{equation}   
		with $k \approx 1.6$ being a dimensionless number.


\section{Washout Analysis}
\label{app:washout}
In this appendix we provide a detailed analysis of
why the three-body decay $\psi \to \chi f\bar f$ and portal interactions do not wash out either the dark-sector or visible-sector 
asymmetries.  



After the heavy scalar doublets $\xi_i$ have decayed and generated the
primordial $L$ and $D$ asymmetries, the remaining dark-sector dynamics 
involves the following processes:
\begin{itemize}
\item the three-body decay and inverse decay
\[
\psi \;\leftrightarrow\; \chi f\bar f, 
\]
\item $Z'$-mediated dark--visible scatterings
\[
\psi f \leftrightarrow \chi f, 
\qquad 
\chi \chi \leftrightarrow f\bar f,
\qquad
\psi\bar\psi \leftrightarrow f\bar f,
\]
\item the Yukawa portal interaction $\lambda\,\psi\,\eta\,\chi$, which is 
irrelevant once $\eta$ becomes Boltzmann suppressed.
\end{itemize}

For washout to occur, two conditions must be satisfied:
(i) the relevant interactions must violate the quantum number that carries 
the asymmetry (here the dark charge $D$ or the lepton number $L$), and 
(ii) the violating processes must be in chemical equilibrium, 
$\Gamma \gtrsim H$, so that the corresponding chemical potentials are 
driven to zero.  As shown below, neither condition is realised in our 
model after the decay epoch of the heavy scalar doublets.


\subsection*{Three body decay and inverse decay $\chi f\bar f \leftrightarrow \psi$}

The three-body decay $\psi\to\chi f\bar f$ is induced by the soft $\mathcal{Z}_2$-breaking operator
$\mu^2\,\eta^\dagger H$.  
Crucially, although this term breaks the discrete parity, it preserves
the dark charge $D$:
\[
D(\psi)=D(\chi),
\qquad
D(f)=D(\bar f)=0.
\]
Therefore the decay merely transfers the primordial $\psi$ asymmetry into the 
stable state $\chi$, and the total
\[
Y_{\Delta D} = Y_{\Delta \psi} + Y_{\Delta \chi}
\]
is conserved exactly.  


Inverse decays $\chi f\bar f \to \psi$ are also $D$-conserving and hence cannot erase $Y_{\Delta D}$.  
Furthermore, even ignoring charge considerations, they are extremely 
suppressed. Using the three-body decay width,
\[
\Gamma_\psi \sim 
\frac{y_{\rm eff}^4}{192\pi^3}\,
\frac{M_\psi^5}{\Lambda^4},
\]
we get the scaling relation,
\[
\boxed{
\Gamma_\Psi \;\simeq\;
1.68\times 10^{-29}\ {\rm GeV}\;
y_{\rm eff}^{4}\,
\left(\frac{M_\psi}{1~{\rm TeV}}\right)^{5}
\left(\frac{10^{10}~{\rm GeV}}{\Lambda}\right)^{4}
}
\]
Even for the most conservative values,  
\[
y_{\rm eff}=1,\quad
M_\psi=1~\mathrm{TeV},\quad
\Lambda=10^{10}~\mathrm{GeV},
\]
one obtains
\[
\Gamma_\psi 
\simeq 1.68\times 10^{-29}~\mathrm{GeV},
\qquad
\frac{\Gamma_\psi}{H(T=M_\psi)} 
\simeq 10^{-13}.
\]
The inverse decay rate is further suppressed by SM thermal number 
densities in a three-body initial state.  
Thus $\Gamma_{\rm inv}/H\ll10^{-13}$ and inverse decays are 
always out of equilibrium.

\subsection*{Portal-mediated scatterings:}

The only remaining processes that could communicate between the
dark and visible sectors are the $Z'$-mediated scatterings,
\[
\psi f \leftrightarrow \chi f,
\qquad
\chi\chi \leftrightarrow f\bar f,
\qquad
\psi\bar\psi\leftrightarrow f\bar f.
\]
These preserve $D$, but if they were fast enough, they could in principle
transfer the dark asymmetry into the visible sector and wash out the 
lepton asymmetry by coupling efficiently to the $L$-violating interactions 
in the $\xi_i$ sector.

In our model the $Z'$ is a light mediator in the MeV range, and the
dark gauge coupling $g'$ is relatively large ($g'\sim 0.1$) in order to
generate the desired self-interacting dark matter cross section.  
However, the potentially dangerous washout processes are not controlled
by $g'$ alone, but by the portal coupling of $Z'$ to the SM.  
This portal is very small in our setup from the requirement of direct search. The kinetic mixing parameter
$\epsilon$ is required to be very small ($\epsilon \lesssim 10^{-8}$), as shown in Fig.~14 of
the paper.  As a result, the effective coupling that enters
SM-dark scatterings is
\[
g'_{\rm eff} \equiv \epsilon\,g' \sim 10^{-9},
\]
which is many orders of magnitude smaller than the dark matter self-interaction coupling.
For relativistic $Z'$-mediated scatterings the thermally averaged cross
section scales as $\langle\sigma v\rangle \sim (g'_{\rm eff})^4/T^2$, while
the number density of SM fermions scales as $n(T)\sim T^3$.  Thus the
scattering rate may be written as
\[
\Gamma_{\rm scatt}(T)
= n(T)\langle\sigma v\rangle
\simeq c_{\rm scatt}\,(g'_{\rm eff})^4\,T,
\]
where $c_{\rm scatt}$ is an $\mathcal{O}(10^{-1})$ coefficient encoding
spin sums, thermal averaging and numerical phase-space factors.  
At the epoch of $\psi$ decay
($T\simeq M_\psi$), we obtain
\[
\frac{\Gamma_{\rm scatt}}{H}
\simeq \frac{c_{\rm scatt}\,M_{\rm Pl}}{1.66\sqrt{g_*}\,M_\psi}\,(g'_{\rm eff})^4
\simeq 1.1\times10^{16}\,(g'_{\rm eff})^4
\left(\frac{10^3~\mathrm{GeV}}{M_\psi}\right).
\]
For $M_\psi$ at TeV scale, with $g'_{\rm eff}\sim 10^{-9}$ (corresponding to $g'\sim 0.1$ and
$\epsilon\sim10^{-8}$), this gives
\[
\frac{\Gamma_{\rm scatt}}{H}\Big|_{T=M_\psi}\sim 10^{-20},
\]
{\it i.e.,} the scattering rate is too slow compared to the Hubble rate. Thus the dark and visible sectors never approach chemical equilibrium. Potentially dangerous washout channels being many orders of magnitude below the
Hubble rate in the allowed parameter region, the asymmetries produced are safely preserved.

\subsection*{Absence of LNV processes after $\xi_i$ decays}

It is to be mentioned that, even if the portal interactions were faster, they still would not wash 
out the lepton asymmetry unless they also communicated with lepton-number violating (LNV) processes.  
But LNV interactions in our model come solely from the decays and inverse decays of
the heavy $\xi_i$ fields.  
Once the temperature drops below $T\ll M_{\xi_i}$ these interactions are
Boltzmann suppressed and effectively decoupled.  
Thus there is no operational LNV channel for any dark-visible process to
couple to, and the visible-sector lepton asymmetry remains protected.
Hence neither the dark-sector asymmetry nor the visible lepton asymmetry 
is affected by late-time decays or scatterings, and no washout occurs 
anywhere in the parameter space relevant for successful asymmetry 
generation.




\end{document}